\documentclass[a4paper,11pt]{article}
\usepackage[utf8]{inputenc}
\DeclareUnicodeCharacter{2009}{\,}
\DeclareUnicodeCharacter{03B1}{\ensuremath{\alpha}}
\DeclareUnicodeCharacter{223C}{\ensuremath{\sim}}
\usepackage[normalem]{ulem}
\usepackage{amsmath}
\usepackage{bm}
\usepackage{subcaption}
\usepackage{jcappub} 
\usepackage{lineno}

\arxivnumber{2508.09112}
\title{\boldmath Tracing large-scale structure morphology with multiwavelength line intensity maps}

\author[a]{Manas Mohit Dosibhatla,}
\author[a]{Suman Majumdar,}
\author[a,b]{Chandra Shekhar Murmu,}
\author[a]{Samit Kumar Pal,}
\author[c]{Saswata Dasgupta,}
\author[d,e]{Satadru Bag,}
\author[a]{Abhirup Datta}

\affiliation[a]{
Department of Astronomy, Astrophysics and Space Engineering,
Indian Institute of Technology Indore, Khandwa Road, Indore - 453552, India
}

\affiliation[b]{
Astrophysics Research Center of the Open University (ARCO) \& Department of Natural Sciences,
The Open University of Israel,
1 University Road, Ra'anana 4353701, Israel
}

\affiliation[c]{
Institute of Astronomy \& Kavli Institute for Cosmology, University of Cambridge,
Madingley Road, Cambridge CB3 0HA, United Kingdom
}

\affiliation[d]{
Physics Department, TUM School of Natural Sciences,
Technical University of Munich,
James-Franck-Stra{\ss}e 1, 85748 Garching, Germany
}

\affiliation[e]{
Max-Planck-Institut f{\"u}r Astrophysik,
Karl-Schwarzschild Stra{\ss}e 1, 85748 Garching, Germany
}

\emailAdd{dosibhatla.mohit@gmail.com}

\abstract{Line intensity mapping (LIM) is an emerging technique for probing the large-scale structure (LSS) in the post-reionisation era. This captures the integrated flux of a particular spectral line emission from multiple sources within a patch of the sky without resolving them. Mapping different galaxy line emissions, such as the HI $21$-cm and CO rotational lines via LIM, can reveal complementary information about the bias with which the line emitters trace the underlying matter distribution and how different astrophysical phenomena affect the clustering pattern of these signals. The stage at which the structures in the ``cosmic web" merge to form a single connected structure is known as the percolation transition. Using mock HI $21$-cm and CO($1-0$) LIM signals in the post-reionisation universe, we explore the connectivity of structures through percolation analysis and compare it with the underlying galaxy distribution. We probe the relative contributions of voids, filaments, and sheets to the galaxy density and line intensity maps using a morphological measure known as the local dimension. The CO($1-0$) map exhibits an increased filamentary behaviour and larger contribution from sheets than the $21$-cm map. We attempt to explain such an emission of the CO($1-0$) line from biased environments. The upcoming SKA-Mid will produce tomographic intensity maps of the $21$-cm signal at $z \lesssim 3$ in Band-1. CO maps can be produced at these redshifts in phase 2 of SKA-Mid, where the frequency coverage is expected to increase up to $\sim 50$ GHz. We present forecasts for the recovery of the local dimensions of these line intensity maps contaminated by thermal noise and line interlopers in SKA-Mid surveys.}

\begin{document}
\maketitle
\flushbottom

 \section{Introduction}
\label{sec:intro}

The large-scale structure (LSS) of the universe takes the form of the so-called ``cosmic web", composed of highly dense nodes that host galaxy clusters, filamentary structures that bridge these nodes, surface-filling structures called sheets, and large underdense regions called cosmic voids \citep{gregory_comaa1367_1978, einasto_structure_1984, zeldovich_giant_1982, bond_how_1996, bharadwaj_evidence_2000, aragon-calvo_multiscale_2010, libeskind_tracing_2018, tojeiro_large_2025}. Such a morphology of the matter distribution has been observed both in cosmological simulations \citep{bond_how_1996, galarraga-espinosa_evolution_2024, zhang_statistical_2024} and observational probes of the LSS, such as galaxy redshift surveys \citep{van_de_weygaert_observations_2008, xia_gravitational_2020, bacon_muse_2021, martin_extensive_2023}. The cosmological model determines the gravitational clustering of dark matter that leads to the web-like structure. Baryonic matter falls into the gravitational potential wells created by dark matter and traces its distribution \citep{loeb_first_2013}. The baryonic tracers (galaxies and intergalactic gas) are also affected by the complex astrophysics of star formation, supernova and AGN feedback, galactic winds, etc. \citep{fabian_observational_2012, tumlinson_circumgalactic_2017, peroux_cosmic_2020, thompson_theory_2024, harrison_observational_2024}. Hence, probing the cosmic web is of great interest because it contains a wealth of information on both cosmology and astrophysics.

The most popular approach to studying the LSS is through galaxy redshift surveys, which rely on detections of individual galaxies with a high signal-to-noise ratio (SNR) and estimating their redshifts via spectroscopic or photometric means [\citealp[and references therein]{schlafly_survey_2023}]. Galaxy surveys have been extensively used to constrain cosmological model parameters \citep{sanchez_cosmological_2006, tegmark_cosmological_2006, blake_wigglez_2011, ivanov_cosmological_2020, des_collaboration_dark_2022,adame_desi_2025}. However, due to the magnitude-limited nature of such surveys, they are biased towards a bright sample of the galaxy population within the survey volume. As a result, they yield clustering and morphology patterns biased towards the brightest sources. Additionally, the accuracy of redshift determination is affected by low SNR and contamination from interloper lines \citep{cagliari_correcting_2025, collaboration_euclid_2025}. 

Line intensity mapping (LIM) is an alternative approach to tracing the LSS. It is based on the idea that to map the clustering of the matter distribution as a whole, it is not necessary to resolve individual gravitationally bound structures (galaxies). LIM involves observing the aggregate emission of a target spectral line from coarse patches of the sky, with the line-of-sight information embedded in the different frequency channels of the instrument, allowing us to construct a tomographic map of the line emission \citep{bharadwaj_using_2001, visbal_measuring_2010, visbal_demonstrating_2011, uzgil_measuring_2014, kovetz_astrophysics_2019, schaan_astrophysics_2021, bernal_line-intensity_2022}. By targeting a specific spectral line emission originating from galaxies, it is possible to map the LSS in a relatively unbiased manner by capturing the flux contribution towards the line emission from faint galaxies as well.

Probing the LSS through intensity maps of multiple spectral lines allows us to obtain complementary information from galaxies, as different lines originate from different regions of galaxies and are affected by different astrophysical phenomena. Observing and cross-correlating lines in separate frequency bands also suppresses the correlation between astrophysical foregrounds and other systematics in the observations, resulting in higher SNRs \citep{murmu_c_2021, cunnington_detecting_2022, berti_21_2024}. Since hydrogen is the most abundant element in the Universe and therefore in galaxies as well, it is suitable to target emissions that can trace it to probe the LSS.  The atomic phase of hydrogen (HI) can be effectively traced by the hyperfine $21$-cm emission, and the molecular phase (H$_2$) by the line-emission arising from the transition between the $J\rightarrow J-1$ rotational levels of the carbon monoxide (CO) molecule \citep{bolatto_co--h2_2013, carilli_cool_2013}, which is the second most abundant molecular species in molecular clouds within galaxies.

Several experiments are underway to probe the HI $21$-cm intensity mapping signal in the post-reionisation universe, and statistical detections and upper limits have been reported by the survey of the ELAIS-N1 field by the uGMRT \citep{chakraborty_first_2021, elahi_towards_2024}, the CHIME collaboration \citep{CHIME_2025_detection, collaboration_detection_2023, collaboration_detection_2023-1}, interferometric surveys of MeerKAT \citep{paul_first_2023, mazumder_hi_2025}, and the single dish MeerKLASS survey \citep{meerklass_collaboration_meerklass_2025, cunnington_detecting_2022} among others. The advent of the Square Kilometre Array Observatory (SKAO) will allow tomographic intensity mapping of the $21$-cm line across different redshifts. SKA-Mid, the mid-frequency component of the SKAO situated in South Africa, will be capable of producing tomographic intensity maps of the $21$-cm line in the redshift range $0 \leq z \lesssim 3$ \citep{santos_cosmology_2015}. Multiple experiments have reported upper limits of the CO LIM power spectrum, such as the CO Power Spectrum Survey (COPSS), Atacama Large Millimetre Array (ALMA), Atacama Compact Array (ACA), Carbon monOxide Mapping Array Project (COMAP), etc. \citep{keating_first_2015, keating_intensity_2020, stutzer_comap_2024}. In this work, we present a morphological analysis of the cosmic web using tomographic intensity maps of the HI $21$-cm line and the $115.27$ GHz CO($1-0$) line simulated by postprocessing IllustrisTNG \citep{nelson_illustristng_2021, pillepich_first_2018, springel_first_2018, nelson_first_2018, naiman_first_2018, marinacci_first_2018} galaxy catalogues. We examine the biases with which these two lines trace the underlying galaxy distribution and differences in the overall morphology of the intensity maps.

The relation between the matter distribution and its observable tracers has been quantified through bias parameters determined using $N$-point correlation functions in the Fourier space \citep{sarkar_modelling_2016, sarkar_modelling_2019, chung_comap_2024}. The power spectrum is the conventional statistic used to quantify the clustering pattern of cosmological fields, but it cannot completely describe the matter distribution in the late universe due to its highly non-Gaussian nature \citep{peebles_large-scale_1980}. Higher-order statistics, e.g., bispectrum, trispectrum \citep{peebles_large-scale_1980, sarkar_modelling_2019, moodley_cross-bispectrum_2023}, marked correlation functions [\citealp[and references therein]{massara_cosmological_2023, kamran_re-markable_2024}], etc., capture more information on such fields. However, they cannot preserve the complete phase information of the signal in the Fourier domain. This leads to information loss on the morphological features of the fields. Therefore, we employ morphological measures directly in real space (image/map domain) to describe the LIM signals. Various image-based statistics, e.g., Minkowski functionals \citep{mecke_robust_1993, bag_shape_2018, pathak_distinguishing_2022, ghara_morphology_2024, bashir_local_2025}, tidal tensor \citep{hahn_properties_2007, foreroromero_dynamical_2009, aycoberry_theoretical_2023}, percolation analysis \citep{shandarin_percolation_1983, klypin_percolation_1993, bharadwaj_evidence_2000, pathak_distinguishing_2022, dasgupta_interpreting_2023, regos_percolation_2024, pal_interpreting_2025}, local dimension \citep{sarkar_local_2009, sarkar_exploring_2012, sarkar_unravelling_2019, pandey_exploring_2020}, etc., have been formulated in the literature to describe cosmological fields.

Percolation analysis \citep{shandarin_percolation_1983, klypin_percolation_1993, bharadwaj_evidence_2000, pathak_distinguishing_2022, dasgupta_interpreting_2023, regos_percolation_2024, pal_interpreting_2025} is the study of the percolation transition in a field, which refers to the onset of large-scale connectivity in the field. It was shown in \citep{bharadwaj_evidence_2000} that the percolation curve of a non-Gaussian and filamentary cosmological field shows an earlier percolation transition than a Poisson random field with the same average density. The percolation transition can therefore be treated as a measure of the degree of filamentarity in a field. In this work, we carry out a percolation analysis of the $21$-cm and CO($1-0$) intensity maps and compare their percolation transition points with each other and also with that of the underlying galaxy mass density field. We then investigate the origin of the different degrees of overall filamentarity of $21$-cm and CO($1-0$) maps by probing the relative abundances of regions lying in filaments, sheets, and volume-filling environments (nodes and voids). This is done using an exponent known as local dimension \citep{sarkar_local_2009, sarkar_exploring_2012, sarkar_unravelling_2019, pandey_exploring_2020}, which determines how the number of points in a sphere centred at a given map cell scales with the radius of the sphere. This measure was originally defined for galaxy samples \citep{sarkar_local_2009}, and we extend it to gridded intensity maps. Through percolation and local dimension analyses, we probe how well $21$-cm and CO($1-0$) line intensity maps trace the underlying galaxy population. We attempt to explain differences in the morphologies of $21$-cm and CO($1-0$) intensity maps in terms of the astrophysical processes that affect the gas responsible for these emissions.

In the proposed phase $2$ of its operation, the highest frequency at which SKA-Mid can observe is expected to be upgraded from the present $\simeq 15$ GHz to $\simeq 50$ GHz \citep{braun_anticipated_2019}. This will open up an opportunity to observe the rotational CO($1-0$) line, having rest frame frequency $115.27$ GHz, originating from redshifts $z \gtrsim 1.3$ using SKA-Mid.\footnote{\href{https://www.skao.int/sites/default/files/documents/d38-ScienceCase_band6_Feb2020.pdf}{SKA1 Beyond 15GHz:
The Science case for Band 6; see Section 6.7.}} This effectively yields a redshift window $1.3 \lesssim z \lesssim 3$ in which the SKA-Mid phase $2$ (SKA2-Mid) will be, in principle, capable of mapping the Universe using both the $21$-cm and CO($1-0$) lines. In \ref{subsec:noise}, we demonstrate that tomographic maps of the two lines with $\sim$Mpc spatial resolutions have the best overall signal-to-noise ratio (SNR) at $z \simeq 1.41$, considering the array assembly $4$ (AA$4$) \citep{seethapuram_sridhar_2025_AA4}
sensitivity estimates of SKA-Mid. Therefore, throughout this article, we present an analysis of the morphology of intensity maps of both of these lines at $z=1.41$. We also provide estimates for the detectability of the percolation statistics and local dimensions in realistic interferometric surveys by SKA-Mid in the presence of thermal noise and line interlopers of the CO($1-0$) line.

This article is organised as follows: Details of the density map, line intensity map, thermal noise, and line interloper simulations are given in section \ref{sec:simulations}; the methodology followed and morphological measures used are described in section \ref{sec:methodology}; results of the morphological analysis are reported in section \ref{sec:results}; conclusions and scope of follow-up work are discussed in section \ref{sec:conclusions}. We follow the $\Lambda$CDM cosmology of \citep{ade_planck_2016}: $h = 0.6774$, $\Omega_m = 0.3089$, $\Omega_b = 0.0486$, $\Omega_\Lambda = 0.6911$. All distances are comoving, and logarithms are base $10$.
\section{Simulations}
\label{sec:simulations}

\subsection{Galaxy catalogue}
We use galaxy catalogues prepared using the subfind algorithm \citep{springel_populating_2001} from the publicly available TNG300-1 box\footnote{\href{https://www.tng-project.org/data/downloads/TNG300-1/}{https://www.tng-project.org/data/downloads/TNG300-1/}} of the cosmological gravo-magnetohydrodynamical simulation suite IllustrisTNG \citep{nelson_illustristng_2021, pillepich_first_2018, springel_first_2018, nelson_first_2018, naiman_first_2018, marinacci_first_2018}, which contains both dark matter and baryonic gas, as the starting point to generate mock line intensity maps. The simulation cube considered here has a comoving side length of $302.6$ Mpc with dark matter and galaxy mass resolutions of $3.98 \times 10^7 \, h^{-1} M_\odot$ and $1.27 \times 10^9 \, h^{-1} M_\odot$, respectively. The MHD equations are solved in a moving mesh grid of average gas cell mass $7.44 \times 10^6 \, h^{-1}M_\odot$. For more details on these simulations, interested readers may refer to \citep{nelson_illustristng_2021, pillepich_first_2018, springel_first_2018, nelson_first_2018, naiman_first_2018, marinacci_first_2018}. We extract the centre of mass coordinates of galaxy positions, total mass ($M$), gas mass ($M_{\rm gas}$), hydrogen mass fraction ($f_{\rm H}$), and instantaneous star formation rate (SFR) of each galaxy in the simulation. The z-axis is chosen to be the line of sight, and the z-component of the peculiar velocity ($v_\parallel$) of each galaxy is used for mapping them in redshift space by implementing redshift space distortion (RSD) \citep{kaiser_clustering_1987, hamilton_linear_1998}. The galaxies are then gridded into two sets of intensity maps for each line emission under consideration, one in real space and the other in redshift space.

\subsection{Density and intensity maps}
We generate $21$-cm and CO($1-0$) intensity maps by postprocessing the galaxy catalogues using the Line Intensity Mapping Tool (\texttt{LIMiT}),\footnote{\href{https://github.com/dmmohit/LIMiT.git}{https://github.com/dmmohit/LIMiT.git}} which is an updated version of the \texttt{LIM simulator} \citep{murmu_c_2021}.\footnote{\href{https://github.com/chandra-001/LIM_simulator.git}{https://github.com/chandra-001/LIM\_simulator.git}} Whereas the \texttt{LIM simulator} tool was limited to the [CII]$_{158\mu \rm m}$ and CO lines, \texttt{LIMiT} contains an HI $21$-cm line emission model from galaxies. This tool takes the positions, $v_\parallel$, $M$, $M_{\rm H} = f_{\rm H}M_{\rm gas}$ and SFR of galaxies as input. For implementing RSD, the galaxy positions are displaced along the line of sight by distances proportional to $v_\parallel$. The redshift space positions ($s$) are related to real space positions ($x$) as
\begin{equation}
    s = x + \frac{1+z}{H(z)}v_\parallel \,.
\end{equation}
\noindent
The galaxy positions are then mapped onto a grid of the desired grid size and resolution. We generate three cosmological fields:
\begin{enumerate}
    \item Galaxy mass density ($\rho_{\rm g}$),
    \item $21$-cm brightness temperature ($T_{21 \rm cm}$),
    \item CO($1-0$) brightness temperature ($T_{\rm CO}$).
\end{enumerate}

\subsubsection{Galaxy mass density ($\rho_{\rm g}$)}
The galaxy mass density map is generated by assigning galaxy mass densities to grid cells. We work with spatial resolutions of the order of hundreds of kpc, which is $1-2$ orders of magnitude larger than the typical size of galaxies. A galaxy will hence effectively contribute its mass and line luminosity only to a single grid cell of the simulated maps. We therefore employ a nearest grid point (NGP) scheme for generating density and intensity maps, where a galaxy contributes its mass/luminosity to a single grid cell.

\subsubsection{21-cm brightness temperature ($T_{21 \rm cm}$)}
To generate the $21$-cm map, the atomic hydrogen fraction is fixed at a constant value $f_{\rm HI} = 0.7$ following \citep{sun_self-consistent_2019}. The HI masses of galaxies, related to their total hydrogen mass $M_{\rm H}$ as
\begin{equation}
    M_{\rm HI} = f_{\rm HI} \times M_{\rm H} \,,
\end{equation}
are then mapped onto a grid. The HI density map thus obtained is converted to a $21$-cm brightness temperature map using equation (36) of \citep{villaescusa-navarro_ingredients_2018}.
 \begin{equation}
    T_{21 \rm cm} (\boldsymbol{x}) = 189\,h\, \frac{H_0 (1+z)^2}{H(z) \rho _c(z)}\, \rho _{\rm HI}(\boldsymbol{x}) \,,
\end{equation}
\noindent
where $h = H_0/(100 \text{ km s}^{-1}\text{Mpc}^{-1})$ is the dimensionless Hubble constant, $H(z)$ and $\rho _c(z)$ are the Hubble parameter and the critical density at redshift $z$ respectively, and $\rho _{\rm HI}(\boldsymbol{x})$ is the HI mass density at a position $\boldsymbol{x}$.

\subsubsection{CO($1-0$) brightness temperature ($T_{\rm CO}$)}
\label{subsubsec:CO_simulations}
In the case of the CO($1-0$) line, the SFRs of galaxies are converted to infrared luminosities ($L_{\rm IR}$) following \citep{jr_star_1998}:
\begin{equation}
    \frac{L_{\rm IR}}{L_\odot} = \left(\frac{\rm SFR}{M_\odot {\rm /yr}}\right) \times 10^{10} \,.
\end{equation}
The $L_{\rm IR}$'s are converted to CO luminosities $L_{\rm CO}$ using the following relations:
\begin{equation}
    \log \left( \frac{L_{\rm CO}^\prime}{\rm K\, km\, s^{-1}\, pc^2} \right) = \frac{1}{\alpha} \left[ \log\left( \frac{L_{\rm IR}}{L_\odot} \right) - \beta \right] \,,
\end{equation}
\begin{equation}
    \frac{L_{\rm CO}}{L_\odot} = 4.9 \times 10^{-5} J^3 \left( \frac{L_{\rm CO}^\prime}{\rm K\, km\, s^{-1}\, pc^2} \right) \,,
\end{equation}
where the quantum number $J$ specifies the rotational energy level of the CO molecule. The CO emission results from a $J \xrightarrow{} J-1$ transition. We follow the parametrisation of \citep{li_connecting_2016} and \citep{keating_intensity_2020}, which uses the values $\alpha = 1.27$, $\beta = -1.00$ estimated in \citep{kamenetzky_co_relations_2016}.

\begin{figure}[htbp]
    \centering
    \includegraphics[width=\textwidth]{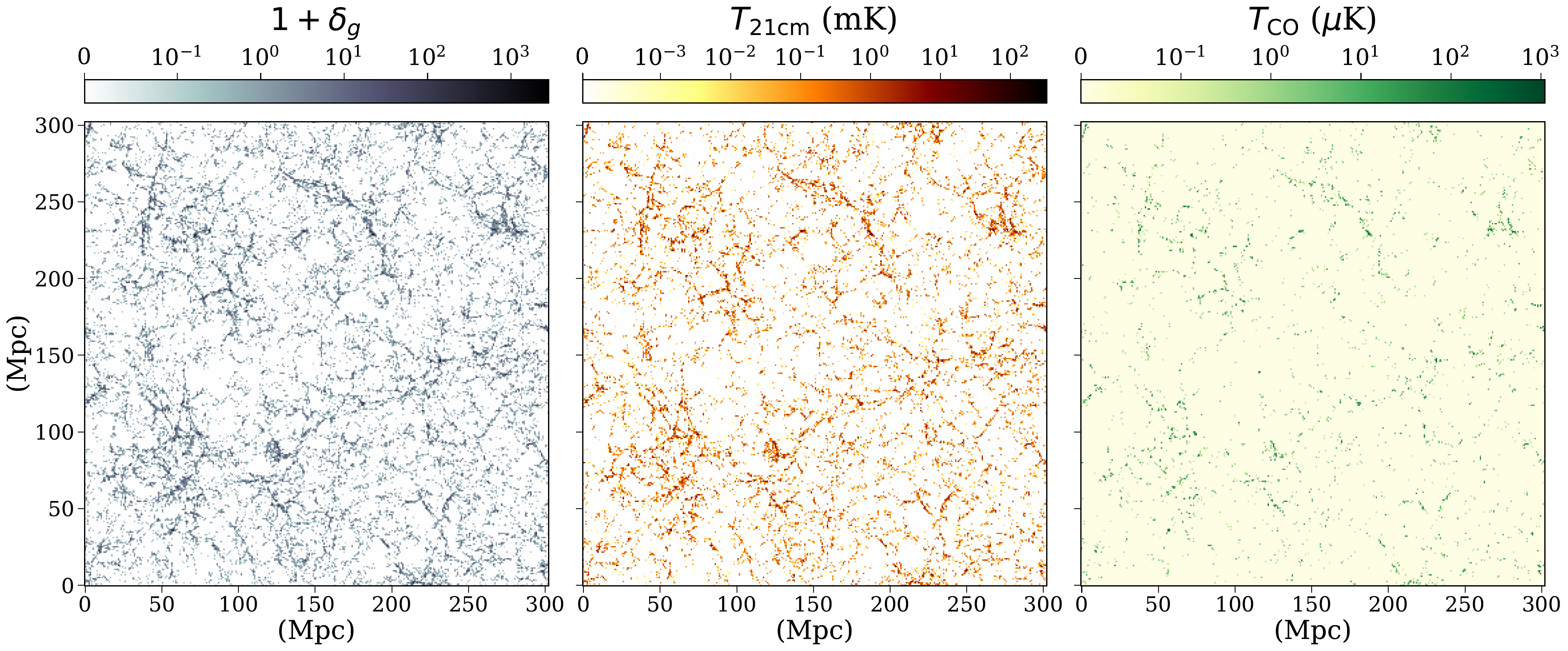}
    \caption{Corresponding slices of the galaxy mass overdensity $1+\delta_{\rm g}$ (left), $21$-cm brightness temperature $T_{21 \rm cm}$ (centre), and CO($1-0$) brightness temperature $T_{\rm CO}$ (right) maps at a spatial resolution $\delta x \simeq 0.87$ Mpc.}
    \label{fig:maps}
\end{figure}

The $L_{\rm CO}$ values are gridded and converted to $T_{\rm CO}$ using equation (1) of \citep{bernal_line-intensity_2022}, which can be rewritten as
\begin{equation}
    T_{\rm CO} = \frac{(1+z)^2 \lambda _{\rm CO}^3}{8\pi k_{\rm B} H(z)}\Bigg( \sum _i L_{\rm CO} (M_i,z) /\Delta V\Bigg)\,,
\end{equation}

where the summation is over galaxies within a grid cell of volume $\Delta V$. Corresponding slices of the $\rho_{\rm g}$, $T_{21 \rm cm}$ and $T_{\rm CO}$ maps at $z=1.41$ are shown in Figure \ref{fig:maps}.

\subsection{Thermal noise simulations}
\label{subsec:noise}

We simulate thermal noise for the SKA-Mid AA4 configuration by assuming the noise to be Gaussian random in the image domain. In practice, thermal noise is Gaussian random in the visibility domain, resulting in pixel-to-pixel correlations in the image domain, which are not considered in the present analysis. The noise maps are hence realisations of a zero-mean Gaussian with noise standard deviation \citep{taylor_synthesis_1999}
\begin{equation}
    \sigma_n = \frac{S_\text{D} \times \text{SEFD}}{\eta_\text{s} \sqrt{\binom{n_\text{ant}}{2} n_\text{pol} \Delta \nu \Delta \tau }} \,,
\end{equation}
where $S_{\rm D} \simeq 2$ is a degradation factor with respect to the natural array sensitivity, $\eta_{\rm s}\simeq0.9$ is the system efficiency, $n_{\rm ant}$ is the number of antennae in the array, $n_{\rm pol}=2$ is the number of polarisations, $\Delta\nu$ and $\Delta\tau$ are the effective spectral resolution and the total integration time respectively. The system equivalent flux density (SEFD) is given by 
\begin{equation}
    \text{SEFD} = 2k_{\rm B}\frac{T_\text{sys}}{A_\text{eff}} \,,
\end{equation}
$k_{\rm B}$, $T_{\rm sys}$, and $A_{\rm eff}$ being the Boltzmann constant, system temperature and effective area of an antenna, respectively. The frequency-dependent ratio $A_\text{eff}/T_\text{sys}$ is taken from the Anticipated SKA1 Science Performance document \citep{braun_anticipated_2019}.\footnote{\href{https://www.skao.int/sites/default/files/documents/SKAO-TEL-0000818-V2_SKA1_Science_Performance.pdf}{https://www.skao.int/sites/default/files/documents/SKAO-TEL-0000818-V2\_SKA1\_Science\_\\Performance.pdf}} We set $n_{\rm ant}=197$, which is the number of antennae in the AA4 configuration of SKA-Mid. The noise $\sigma$'s are estimated in Jy/beam units and are converted to brightness temperature units ($\sigma _T$) using the Rayleigh Jeans approximation
\begin{equation}
    \sigma _T = \frac{\lambda ^2}{2k_{\rm B}\Omega}\sigma_n \,,
\end{equation}
$\lambda$ being the redshifted wavelength of the line. The $\Omega$ is the solid angle subtended by the $\delta x \delta y$ resolution element at the observer's location. It is calculated as $\Omega = \pi \theta ^2/(4 \ln 2)$, where $\theta$ is the corresponding plane angle. In our simulations, the grid cells are cubical, and thus, we assume that the resolution of the sky plane ($\delta x = \delta y$) is the same as the resolution along the line of sight ($\delta$z). The $\delta$z is computed from the change in redshift $\Delta z$ along the line of sight corresponding to the spectral resolution $\Delta\nu$, 
\begin{equation}
    \Delta z = (1+z)^2\bigg(\frac{\Delta\nu}{\nu_0}\bigg) \,.
\end{equation}
\noindent
Detecting filamentary features in the intensity maps requires a high spatial resolution, which means a small $\Delta z$. This, in turn, yields a small $\Delta\nu$ and increases the noise level. An optimal value for the spectral and, hence, spatial resolution must be chosen for our maps to strike a balance between a moderate SNR ($\gtrsim 1$) and high spatial resolution. We find the optimal value to be $\Delta\nu = 107.52 \,\rm kHz$, which is the effective spectral resolution after averaging over eight SKA-Mid frequency channels.\footnote{\href{https://sensitivity-calculator.skao.int/}{https://sensitivity-calculator.skao.int/}} Even at this resolution, thermal noise remains dominant over the signal at most redshifts.

\begin{figure}[htbp]
    \centering
    \includegraphics[width=0.6\linewidth]{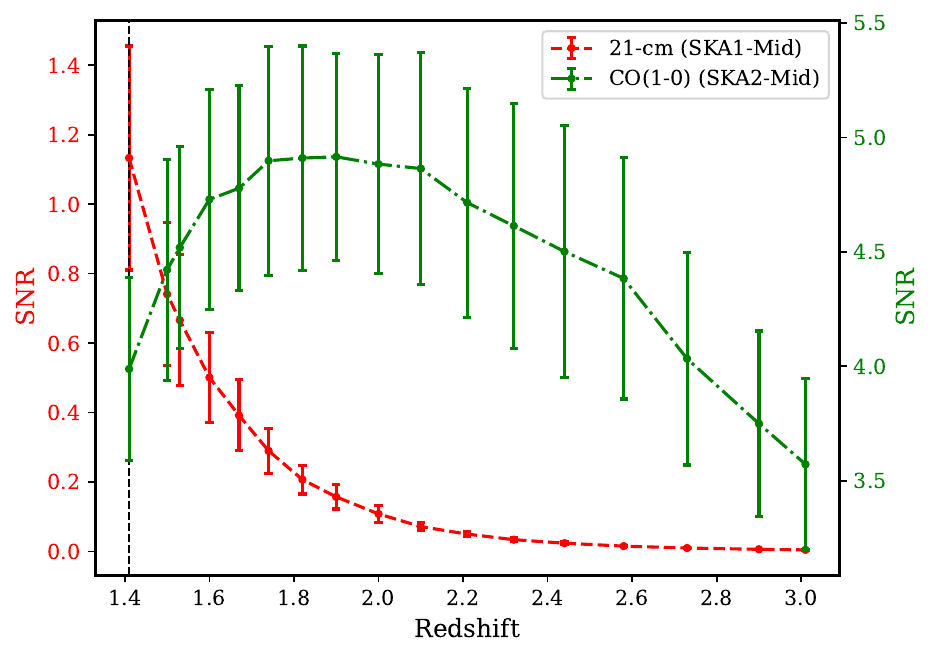}
    \caption{Signal-to-noise ratios (SNR) of the $21$-cm and CO($1-0$) brightness temperature fluctuations at spatial resolution $\delta x \simeq 1$ Mpc in the overlapping redshift range where SKA1-Mid and SKA2-Mid can observe the $21$-cm and CO($1-0$) LIM signals, respectively. The noise estimates are made assuming Gaussian random thermal noise for $5000$ hours per pointing of SKA1-Mid observations and $100$ hours per pointing of SKA2-Mid observations for the $21$-cm and CO($1-0$) signals, respectively. The black dashed line corresponds to $z=1.41$, the redshift where both the $21$-cm and CO($1-0$) brightness temperature fluctuations have a significant SNR. The $1\sigma$ error bars denote standard deviation in SNR across $8 \, (150 \, {\rm Mpc})^3$ 21-cm subcubes and $27 \, (100 \, {\rm Mpc})^3$ CO($1-0$) subcubes, corresponding to realistic SKA-Mid surveys, due to cosmic variance in signal fluctuations.}
    \label{fig:SNR}
\end{figure}

The variation of the SNR with redshift at $\delta x \simeq 1$ Mpc is shown in Figure \ref{fig:SNR}. The SNR is defined as the ratio of the standard deviation of the brightness temperature fluctuations of the signal to that of the noise (SNR $= \, \sigma_{\rm signal}/\sigma_T$). The redshift range corresponds to the set of redshifts where SKA1-Mid can observe the $21$-cm signal, and SKA2-Mid can observe the CO($1-0$) signal. The redshift values are those at which the IllustrisTNG galaxy catalogues are available. We estimate thermal noise for $5000$ hours of SKA1-Mid observation per pointing for the $21$-cm signal and $100$ hours of SKA2-Mid observation per pointing for the CO($1-0$) signal. The $21$-cm maps are highly noise-dominated at almost all redshifts, and the SNR crosses $1$ at $z \simeq 1.5$. In the rest of the article, we carry out our morphological analysis at $z=1.41$, where the $21$-cm signal fluctuations have SNR $>1$ and the CO($1-0$) fluctuations have a significant SNR ($\simeq 4$) as well. The spectral resolution element $\Delta\nu = 107.52 \,\rm kHz$ corresponds to a comoving spatial resolution $\simeq 0.87$ Mpc at $z=1.41$.

A $300 \, {\rm Mpc} \times 300 \, {\rm Mpc}$ patch at $z=1.41$ spans $\simeq 16 \, {\rm deg}^2$ on the sky. The field of view ($\simeq 1.22 \, \lambda / D_{\rm dish}$) of SKA-Mid is limited to $5.56 \, {\rm deg}^2$ for the 21-cm line, and $\simeq 92 \, {\rm arcmin}^2$ with a phased array feed configuration \citep{Hotan_ASKAP_2021} (see section \ref{sec:conclusions}) for the CO($1-0$) line, assuming a constant $D_{\rm dish}$ of $15$m. This would only allow sub-surveys spanning a fraction of the patch within a feasible observation time. We consider $4 \, {\rm deg}^2 \, (150 \, {\rm Mpc} \times 150 \, {\rm Mpc})$ and $1.77 \, {\rm deg}^2 \, (100 \, {\rm Mpc} \times 100 \, {\rm Mpc})$ sub-surveys for 21-cm and CO($1-0$) maps respectively. CO($1-0$) maps formed from 70 non-overlapping pointings must be mosaicked to obtain this sub-survey. We divide the $\simeq (300 \, {\rm Mpc})^3$ LIM cubes into $8 \, (150 \, {\rm Mpc})^3$ 21-cm subcubes, and $27 \, (100 \, {\rm Mpc})^3$ CO($1-0$) subcubes.

The cosmic variance in the SNR of the maps due to the variation of the signal fluctuations among the subcubes at a constant noise level corresponding to 5000 hours and 100 hours per pointing for 21-cm and CO($1-0$) maps is given by $1\sigma$ error bars in Figure \ref{fig:SNR}. The high redshift end is highly noise-dominated for 21-cm maps, and cosmic variance in signal fluctuations is found to have a negligible effect on the SNR with increasing redshift. At $z=1.41$, the SNR ranges between $\sim 0.8-1.5$, and the noise can be suppressed following the methods discussed in \ref{subsec:locdim}. Because CO($1-0$) maps have a much higher overall SNR, cosmic variance is not expected to significantly affect detectability.

\subsection{Line interloper modelling}

Higher $J$ lines in the rotational ladder of the CO molecule act as interlopers to the CO($1-0$) line. For the CO($1-0$) line at $z=1.41$, some of the line interlopers are: CO($2-1$) at $z=3.82$, CO($3-2$) at $z=6.23$, and CO($4-3$) at $z=8.64$. The first two interlopers out of these are expected to contribute the most to the observed signal, as the power spectrum of CO($4-3$) at $z \simeq 8.6$ lies two orders of magnitude below that of the target CO($1-0$) signal.

As the IllustrisTNG galaxy catalogues are constructed from snapshots of the same evolving coeval box, the structures would correspond to the same point in space at different cosmic times. This would make interloper signals modelled from coeval boxes at different redshifts highly correlated. To decrease the statistical dependence of the target and interloper signals, we do the following:
\begin{enumerate}
    \item Simulate three different CO intensity map lightcones: CO($1-0$) centred at the target $z=1.41$, CO($2-1$) centred at $z=3.82$ and CO($3-2$) centred at $z=6.23$.
    \item Rotate the CO($2-1$) lightcone by $90$ degrees and the CO($3-2$) lightcone by 180 degrees about the line of sight.
\end{enumerate}

To construct the lightcones, the following coeval intensity maps are simulated following the prescription in \ref{subsubsec:CO_simulations} using the values of $\alpha$ and $\beta$ as given in Table 4 of \citep{kamenetzky_co_relations_2016}.
\begin{itemize}
    \item CO($1-0$) maps at $z=1.30,\,1.36,\,1.41,\,{\rm and}\,1.50$,
    \item CO($2-1$) maps at $z=3.49,\,3.71,\,4.01,\,{\rm and}\,4.18$,
    \item CO($3-2$) maps at $z=5.85,\,6.01,\,6.49,\,{\rm and}\,7.01$.
\end{itemize}

The lightcones are finally constructed by combining the coeval intensity maps in real space and using the \texttt{LICMAGS}\footnote{\href{https://github.com/chandra-001/LICMAGS.git}{https://github.com/chandra-001/LICMAGS.git}} code that follows the steps described in \citep{murmu_c_2021, datta_light-cone_2012} using the Steffen interpolation scheme \citep{steffen_1990_interpolation}. The redshift slices in the lightcones are equispaced in comoving distance by $0.87$ Mpc, as for the coeval maps. We cut the lightcones so that they have the same dimensions along the sky plane and the line of sight for the sake of simplicity.

\section{Morphological measures}
\label{sec:methodology}

The clustering pattern of the cosmic LSS has traditionally been quantified by the $2$-point correlation function (2PCF) and its Fourier conjugate --- the power spectrum \citep{peebles_large-scale_1980}. However, both of these statistics capture incomplete information on clustering, which becomes increasingly non-Gaussian with cosmic time due to the gravitational clustering of matter. To provide a complete description of the large-scale matter distribution, the infinite hierarchy of $N$-point correlation functions must be computed. This comes with its own computational challenges, as they may not be adequately sampled in cosmological simulations of finite volume with increasing order $N$. The three-point correlation function (bispectrum) [\citealp[and references therein]{peebles_large-scale_1980, sarkar_modelling_2019, moodley_cross-bispectrum_2023}] and marked correlation functions [\citealp[and references therein]{massara_cosmological_2023, kamran_re-markable_2024}] have been studied extensively in the literature in this context, but they lose information on the phase correlations, which is vital for quantifying the clustering pattern of the observed structure.

Alternatively, cosmic structure can be studied directly in the image domain using methods such as Minkowski functionals \citep{mecke_robust_1993, bag_shape_2018, pathak_distinguishing_2022, ghara_morphology_2024, bashir_local_2025}, percolation analysis \citep{shandarin_percolation_1983, klypin_percolation_1993, bharadwaj_evidence_2000, pathak_distinguishing_2022, dasgupta_interpreting_2023, regos_percolation_2024, pal_interpreting_2025}, wavelet transforms \citep{chung_exploration_2022, greig_exploring_2022}, etc.
It is possible for a Gaussian and a non-Gaussian field to have an identical power spectrum, in which case they cannot be distinguished using the 2PCF. However, such scenarios can be distinguished via percolation analysis \citep{bharadwaj_evidence_2000}. This approach involves coarse-graining a gridded field or increasing the linking length of a point distribution until a single contiguous cluster is formed between the two ends of the box. At this stage, the field is said to have percolated, i.e., a pathway would have formed between its ends. Extending this argument with periodic boundary conditions, the cluster would have spanned the universe. This percolation behaviour will be quantified in \ref{subsec:percolation}.

The LSS observed in both galaxy surveys and $N$-body simulations takes the form of the ``cosmic web", which is composed of filaments, sheets, nodes, and voids. Several ways of characterising these building blocks have been proposed in the literature \citep{sarkar_local_2009, hahn_properties_2007, foreroromero_dynamical_2009, hoffman_kinematic_2012, aycoberry_theoretical_2023, sousbie_disperse_2013, platen_cosmic_2007, neyrinck_zobov_2008}. In this work, we employ the simple yet effective quantity, local dimension \citep{sarkar_local_2009, sarkar_exploring_2012, sarkar_unravelling_2019, pandey_exploring_2020}, to determine the kind of environment a particular location in an intensity map lies in as described in \ref{subsec:locdim}.

\subsection{Percolation analysis}
\label{subsec:percolation}

The fraction of cluster volume occupied by the largest cluster in a map is known as the largest cluster statistic (LCS) \citep{klypin_percolation_1993, bharadwaj_evidence_2000, bag_shape_2018, pathak_distinguishing_2022, dasgupta_interpreting_2023, pal_interpreting_2025}. In the case of a gridded field, a cluster may be defined as a connected set of cells with intensity $I$ above or below a threshold value $I_0$. We are interested in overdensities in the galaxy density and line intensity maps as they trace the matter distribution with a characteristic bias. Since all bright (nonzero) cells host galaxies, we fix the threshold at $I_0 = 0$, without considering the noise bias and variance that will be present in the maps. Starting from a bright cell, a cluster is formed using the friends of friends (FoF) algorithm until the cell intensity falls below the threshold value ($I \leq I_0$). The largest of all clusters is identified, and LCS is computed using the \texttt{SURFGEN2} code \citep{bag_shape_2018, bag_studying_2019, sheth_measuring_2003}.
\begin{equation}
    \text{LCS} = \frac{\text{Volume of the largest cluster}}{\text{Total volume occupied by clusters}} \,.
\end{equation}
Another quantity of interest is the filling factor (FF), which is the fraction of the simulation volume occupied by clusters.
\begin{equation}
    \text  {FF} = \frac{\text{Total volume occupied by clusters}}{\text{Volume of the simulation box}} \,.
\end{equation}

\begin{figure}[htbp]
    \centering
    \includegraphics[width=\textwidth]{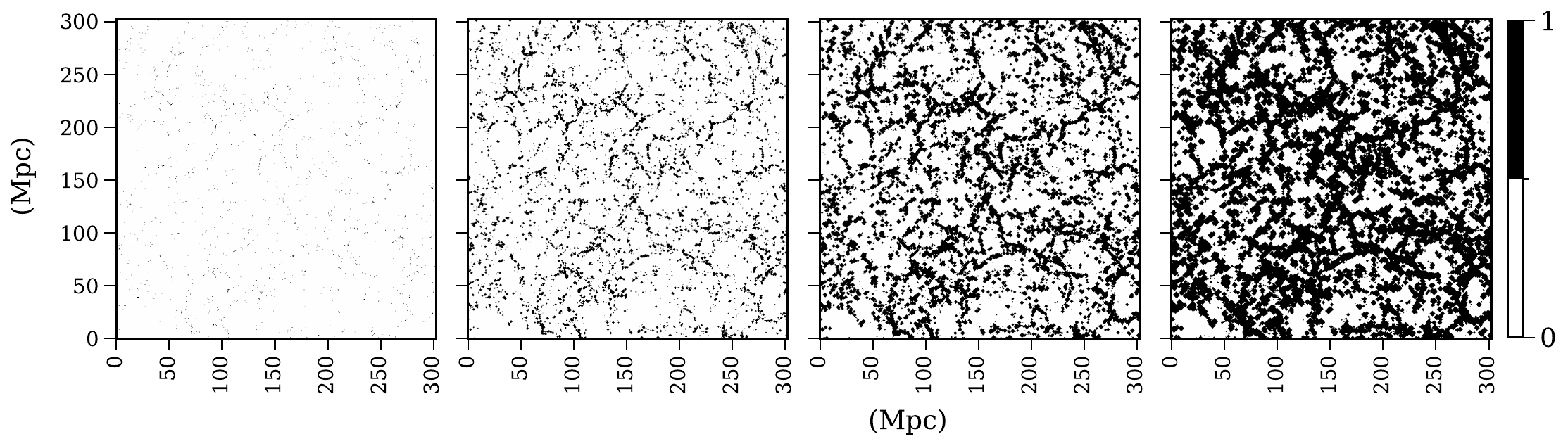}
    \caption{Illustration of the iterative coarse-graining scheme used for percolation analysis. The four panels show the same slice of a map without coarse-graining (first from left) and after 2, 4, and 6 iterations in order. On each iteration, the cells sharing a face with a bright cell are marked as bright. The maps are binary, with the dark cells marked by zeros and the bright cells by ones.}
    \label{fig:coarse-grain}
\end{figure}

We investigate how the LCS varies with FF by iteratively coarse-graining the maps. At every iteration, the cells sharing a face with a bright cell are identified and marked as bright \citep{bharadwaj_evidence_2000} as shown in Figure \ref{fig:coarse-grain}. The FF and LCS of a map are computed after every iteration. The spatial resolution of our maps may not necessarily be adequate to reveal the connectivity of the structure. A cluster of galaxies may appear as a set of disconnected regions due to a small grid spacing. The above coarse-graining scheme allows the growth of structure and reveals its connectivity.

Coarse-graining a map increases its FF, and since it generally leads to the merging of different clusters, the LCS also changes. It has been established that at a certain point, characteristic of the field in question, the LCS shoots up to $\sim1$ for a small change in the FF (see discussions in \citep{shandarin_percolation_1983, bharadwaj_evidence_2000}). At this stage, the field is said to have percolated. We investigate this percolation behaviour for the galaxy mass density, $21$-cm, and CO($1-0$) maps by tracking the variation in LCS with FF.

\subsection{Local dimension}
\label{subsec:locdim}

Given a bright cell in a map, it can be determined whether it lies in a filamentary, sheet-like or volume-filling (nodes and voids) environment using its local dimension defined in \citep{sarkar_local_2009}. This was originally defined for a discrete point distribution, and we extend this to gridded fields (density and line intensity maps).\footnote{\href{https://github.com/dmmohit/local_dimension.git}{https://github.com/dmmohit/local\_dimension.git}} First, we randomly sample a given number of cells out of bright cells having $I > I_0$. For a given value of $I_0$, we count the number of bright cells $N(R)$ in a sphere of radius $R$ centred at a given bright cell and examine how it scales with $R$. The radius $R$ is varied from $R_{\rm min}$ to $R_{\rm max}$ in steps of size $\Delta R$. We make sure that the sampled centres are at least a distance $R_{\rm max}$ apart from the faces of the box, so that periodic boundary conditions need not be applied. Then, $N(R)$ is fitted with a power law
\begin{equation}
    N(R) = AR^D \,,
\end{equation}
where $A$ is a normalisation constant and $D$, which determines the scaling behaviour, is called the local dimension of the bright cell for the length scale $R_{\rm min}$ to $R_{\rm max}$. As in the percolation analysis, we set $I_0 = 0$. Later in this work, we examine the impact of thermal noise and line interlopers on the local dimension analysis.

The exponent $D$ is expected to take the values $1$, $2$, and $3$ for cells in straight filaments, sheets, and volume-filling environments (nodes and voids), respectively. However, fractional values are allowed for intermediate environments between any two types mentioned above \citep{sarkar_local_2009}. We impose the condition $\chi _\nu ^2 \leq 1.2$ for the power law fit to ensure both goodness of fit and a large number of cells where the fits converge. If this condition is satisfied, the local dimension is defined for the given cell and the cell is said to be classifiable. The fraction of cells where $D$ can be defined depends on the length scale. The morphology of a density/intensity map is characterised by plotting the distribution of the $D$-values in a subset of cells sampled from the map. As in \citep{sarkar_unravelling_2019}, we classify the $D$-values into bins of sizes $0.5$, centred at multiples of $0.5$ between and including $1.0$ and $3.0$. The notation used for the different $D$ bins is shown in Table \ref{tab:notation}. The distribution function is defined as
\begin{equation}
    P(D) = \frac{\text{Number of centres in $D$-bin}}{\text{Bin width} \times \text{Total number of centres where $D$ is defined}} \, .
\end{equation}

\begin{table}
    \centering
    \begin{tabular}{|c||c|c|c|c|c|c|}
        \hline
        $D\in$ & $(0.75, 1.25]$ & $(1.25, 1.75]$ & $(1.75, 2.25]$ & $(2.25, 2.75]$ & $(2.75, 3.25]$ \\
        \hline
        Class & C$1$ & I$1$ & C$2$ & I$2$ & C$3$ \\
        \hline
    \end{tabular}
    \caption{Notation used for the different local dimension bins.}
    \label{tab:notation}
\end{table}

Due to a low SNR, recovery of the local dimensions from noisy maps is challenging. As noise is random and uncorrelated, it is expected to wash out the information of the cosmic web structure, and a noise-dominated map would yield $D \simeq 3$ everywhere. To recover the LSS information, we need to either go to the signal-dominated regime or de-noise the maps by smoothing them with a suitable kernel. We describe some approaches to recover the LSS information from noisy intensity maps below:

\begin{itemize}
    \item \textbf{$n\sigma$ thresholding:} The thermal noise in our simulations is Gaussian random, while the signal has a highly skewed distribution. Therefore, on applying a threshold intensity that is some integer $n$ times the noise $\sigma$, we can expect that the cells above the threshold will have a larger signal component. Therefore, to obtain the local dimension distribution of a noisy map, we sample cells randomly out of all the cells that have intensity above $n\sigma$, and count $N(R)$ by including cells with intensity $> \, n\sigma$.

    \item \textbf{De-noising by smoothing:} We find cases with low SNR in section \ref{sec:results} where the structure is washed out even above $1\sigma$ and $2\sigma$ thresholds. In such cases, we smooth the noisy map with a Gaussian smoothing kernel of size ($R_{\rm s}$) that is a multiple of the spatial resolution. As the noise has zero mean and is uncorrelated across grid cells, the smoothed noise component tends to zero for a sufficiently large smoothing kernel size. The signal component survives, even though its magnitude decreases, because the signal is spatially correlated due to the LSS.
    
\end{itemize}

\noindent
We follow a combination of the above two methods to recover the local dimension distribution from noisy maps.
\section{Results}
\label{sec:results}

\subsection{Percolation analysis}

\begin{figure}[htbp]
    \centering
    \includegraphics[width=0.85\textwidth]{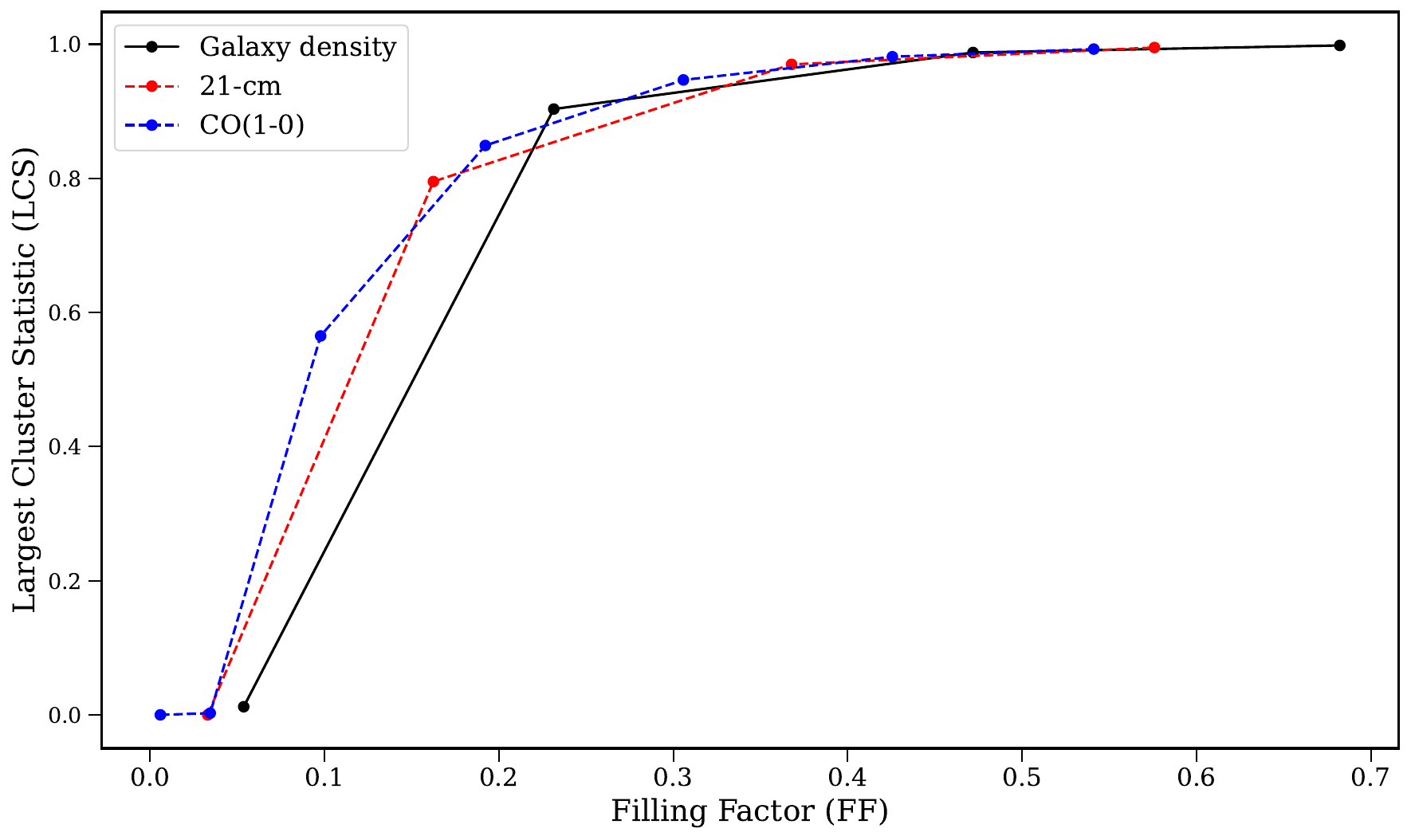}
    \caption{Percolation curves for the galaxy mass density (solid black), $21$-cm (dashed red), and CO($1-0$) (dashed blue) maps at spatial resolution $\delta x = 0.5$ Mpc.
    \label{fig:percolation}}
\end{figure}

Figure \ref{fig:percolation} shows the percolation curves of the galaxy mass density, $21$-cm, and CO($1-0$) maps. The maps have a spatial resolution of $\delta x \simeq 0.5$ Mpc, as it allows low FFs initially such that the percolation transition is visible. The different points on the curve result from the FF and LCS computed after every successive iteration in the coarse-graining scheme. CO($1-0$) emission originates from a subset of galaxies that are star-forming, leading to fewer bright cells in the CO($1-0$) map. Therefore, initially, it has the lowest FF. However, as we iteratively coarse-grain the maps, the CO($1-0$) map percolates faster than the underlying galaxy density map, and the percolation curve for the $21$-cm map lies in between the two. This might indicate that relative to the $21$-cm map, the CO($1-0$) map has a more filamentary morphology, which in turn exhibits a more filamentary morphology than the galaxy density map. As both the FF and the LCS are computed after each step of coarse-graining, the former cannot be independently varied, and the LCS cannot be compared at the same FF values for the three maps for an even comparison. Therefore, the percolation analysis is an inconclusive indicator of the overall filamentarity of the maps. Regardless of their exact nature, the $21$-cm and CO($1-0$) maps having different percolation curves indicate that the emission regions have different connectivity patterns, and originate from biased locations in the cosmic web.

\subsection{Local dimension}

The distribution of local dimension values of the map cells shows the contribution of filaments, sheets and voids to the galaxy densities and line emissions. 

\subsubsection{Uncontaminated signal maps}
\begin{figure}[htbp]
    \centering
    \includegraphics[width=0.65\linewidth]{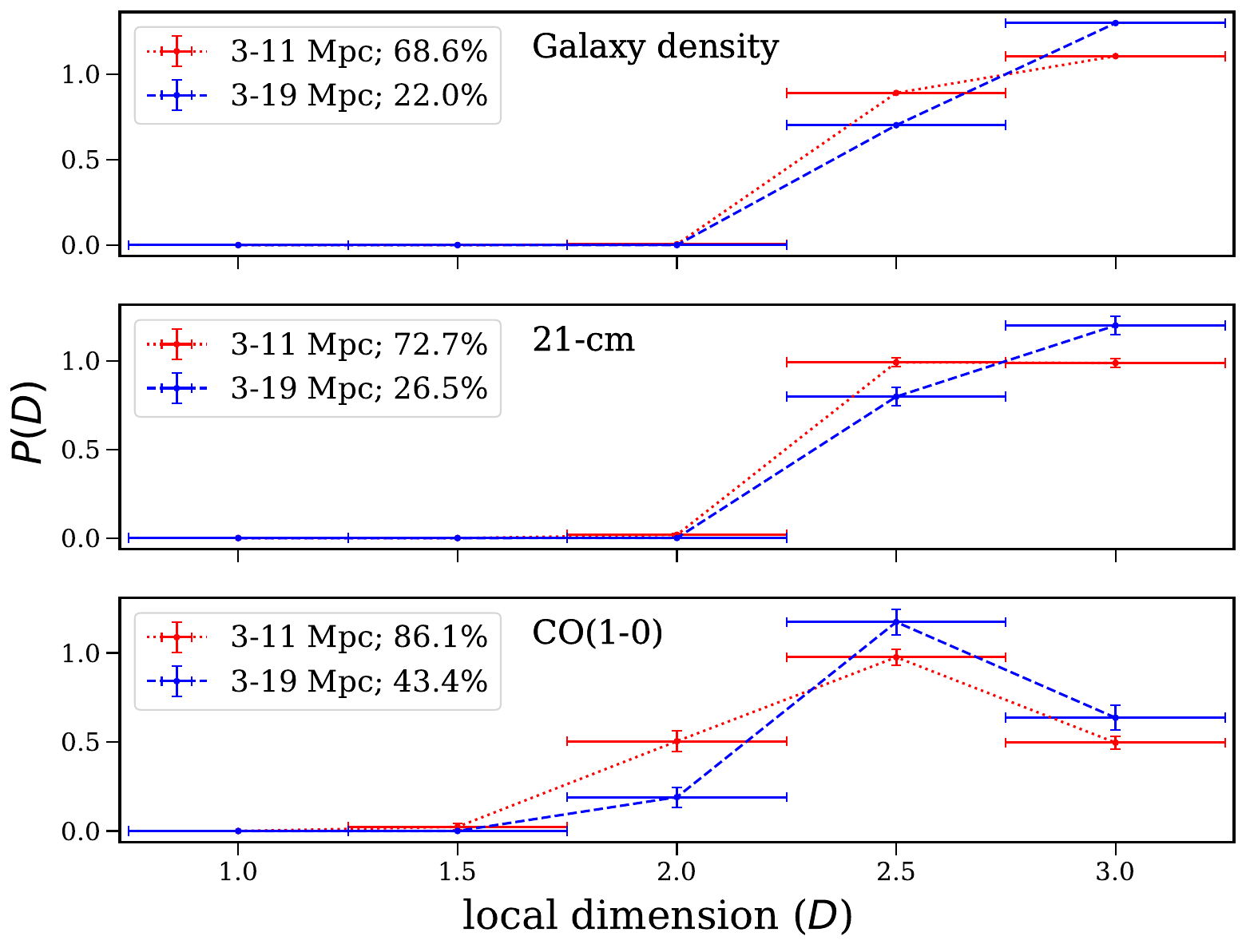}
    \caption{Distribution of local dimensions of classifiable cells out of $10^5$ randomly sampled bright cells in the galaxy mass density (top), $21$-cm (middle), and CO($1-0$) (bottom) maps simulated with galaxy positions in real space. The $D$-values are binned into intervals of size 0.5. The different curves correspond to different length scales specified by the $R_{\rm min}$ and $R_{\rm max}$ values. The percentage of classifiable cells is mentioned alongside the length scale. The $1\sigma$ error bars denote standard deviation in $P(D)$ across $8 \, (150 \, {\rm Mpc})^3$ 21-cm subcubes and $27 \, (100 \, {\rm Mpc})^3$ CO($1-0$) subcubes, corresponding to realistic SKA-Mid survey areas of $4 \, {\rm deg}^2$ and $1.77 \, {\rm deg}^2$ respectively for $10^4$ randomly sampled bright cells.}
    \label{fig:locdim_signal_real}
\end{figure}

Figure \ref{fig:locdim_signal_real} shows the distribution of local dimensions of $10^5$ cells randomly sampled out of the bright cells in the galaxy mass density, $21$-cm, and CO($1-0$) maps. The spatial resolution of the maps is set to $0.87$ Mpc to achieve a balance between a fine resolution to be able to resolve small-scale features, and a coarse enough resolution to reduce thermal noise. As the grid cells are cubical, the spatial resolution in the sky plane is coupled with the spectral resolution (spatial resolution along the line of sight). Unless specified otherwise, all subsequent maps will have a spatial resolution of $0.87$ Mpc. The $R_{\rm min}$, $R_{\rm max}$ and $\Delta R$ values are chosen such that the number of intermediate radii taken for the local dimension fits is equal for each length scale. This is done to maintain consistency and ensure that any difference in $D$ between the length scales results from an actual difference in morphology. We note that emissions from the C3 class (see Table \ref{tab:notation}) originate from voids, as the length scales chosen are larger than the typical sizes of nodes. 

\begin{figure}[htbp]
\centering
\includegraphics[width=0.7\textwidth]{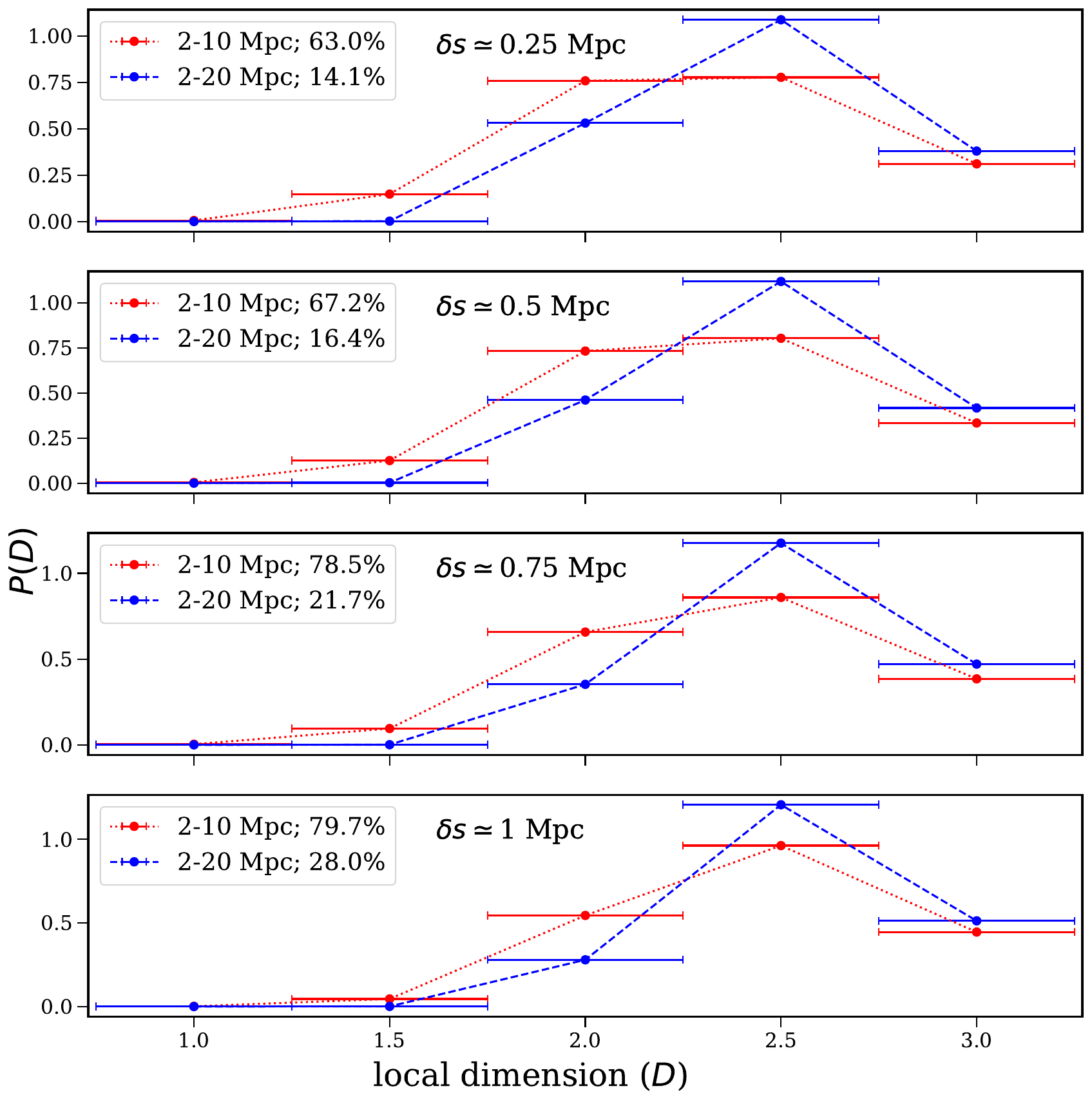}
\caption{Comparison of the distributions of local dimensions of CO($1-0$) maps at spatial resolutions of $\delta s \simeq 0.25,\, 0.5,\, 0.75, {\rm and \,} 1$ Mpc. The length scales are chosen such that they yield similar numbers of $R$ values between $R_{\rm min}$ and $R_{\rm max}$ at each length scale, and $R_{\rm min}$ and $R_{\rm max}$ themselves remain approximately the same for an even comparison.}
\label{fig:resolution_comparison}
\end{figure}

In each map, the contribution from higher $D$-values increases with an increase in length scale. This is expected because as we go to larger length scales, we keep approaching the homogeneity scale where the LSS becomes homogeneous ($D \simeq 3$). For both the galaxy density and the $21$-cm maps, the contribution is the highest from class C3 across length scales, decreasing monotonically with decreasing $D$. The contribution from C1 and I1 is negligible at the map resolution. The comparison between different spatial resolutions is shown in Figure \ref{fig:resolution_comparison} later in this section. However, in the case of the CO($1-0$) map, the contribution from C3 is suppressed and the distribution peaks at I2. There is also an increase in the fraction of centres in I1 and C2.

The percentage of classifiable cells decreases with increasing length scale, because the neighbourhood of a cell is less likely to maintain a similar morphology throughout a radius range as $R_{\rm max}$ goes up, which is a necessary condition for the fits to converge. Overall, the CO($1-0$) map shows a higher degree of convergence of fits than the others. This might be a consequence of most of the CO($1-0$) emission originating from cosmic web galaxies (galaxies lying in filaments, sheets and nodes), where the environments are expected to maintain a similar morphology over large scales. On the contrary, the $21$-cm signal traces the underlying galaxy distribution more closely, and the fits might not converge in a large fraction of cells lying in irregular environments.

The cosmic variance in the distribution of local dimensions in the observable maps, i.e., the 21-cm and CO($1-0$) intensity maps, is denoted by $1\sigma$ error bars in Figure \ref{fig:locdim_signal_real}. This is estimated as the standard deviation in the $P(D)$ values in each $D$-bin across $8 \, (150 \, {\rm Mpc})^3$ 21-cm subcubes and $27 \, (100 \, {\rm Mpc})^3$ CO($1-0$) subcubes, corresponding to SKA-Mid survey areas as detailed in sections \ref{subsec:noise} and \ref{sec:conclusions}. We only find a narrow spread in the local dimension distributions about the morphology of the larger cosmological box at the $1\sigma$ level.

\begin{figure}
    \centering
    \includegraphics[width=0.65\linewidth]{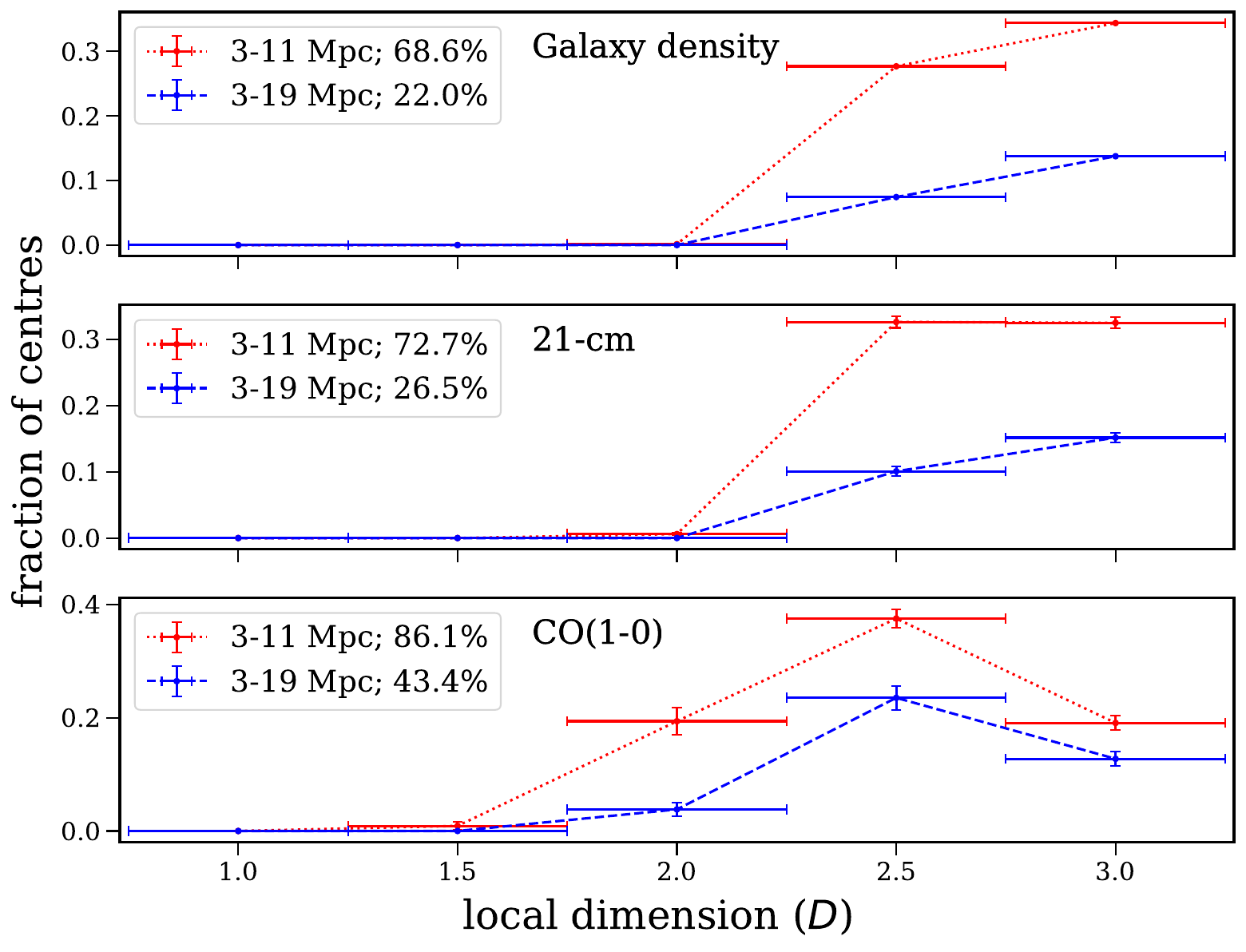}
    \caption{Same as Figure \ref{fig:locdim_signal_real}, normalised with the number of centres sampled instead of the number of classifiable centres. This gives the fraction of cells lying in each bin out of the number of bright cells sampled.}
    \label{fig:locdim_signal_real_fracs}
\end{figure}

Figure \ref{fig:locdim_signal_real_fracs} shows the fractions of the same $10^5$ cells lying in the different $D$-bins. The fractions in different $D$-bins do not add up to $1$ for either length scale because many fits converge to unphysical values above the C3 bin, which have not been shown. A difference between the distributions for the galaxy density and $21$-cm maps, which was not noticeable in Figure \ref{fig:locdim_signal_real}, shows up, particularly in the $3-11$ Mpc length scale. The fraction of cells in C3 remains the same, but the $21$-cm map shows an increased contribution from I2 compared to the galaxy density map. In contrast, the CO($1-0$) map shows reduced contribution from C3 and a significant increase at C2 and I2. The increase in C2 centres comes at the expense of a reduced contribution from C3 centres, as the contribution from I2 does not decrease but instead increases. As in Figure \ref{fig:locdim_signal_real}, we find the $1\sigma$ errors in $P(D)$ to be small compared to the $P(D)$ estimated from the larger box. All subsequent local dimension distributions will be obtained from $10^5$ randomly sampled bright cells unless specified otherwise.

\begin{figure}[htbp]
    \centering
    \includegraphics[width=0.65\linewidth]{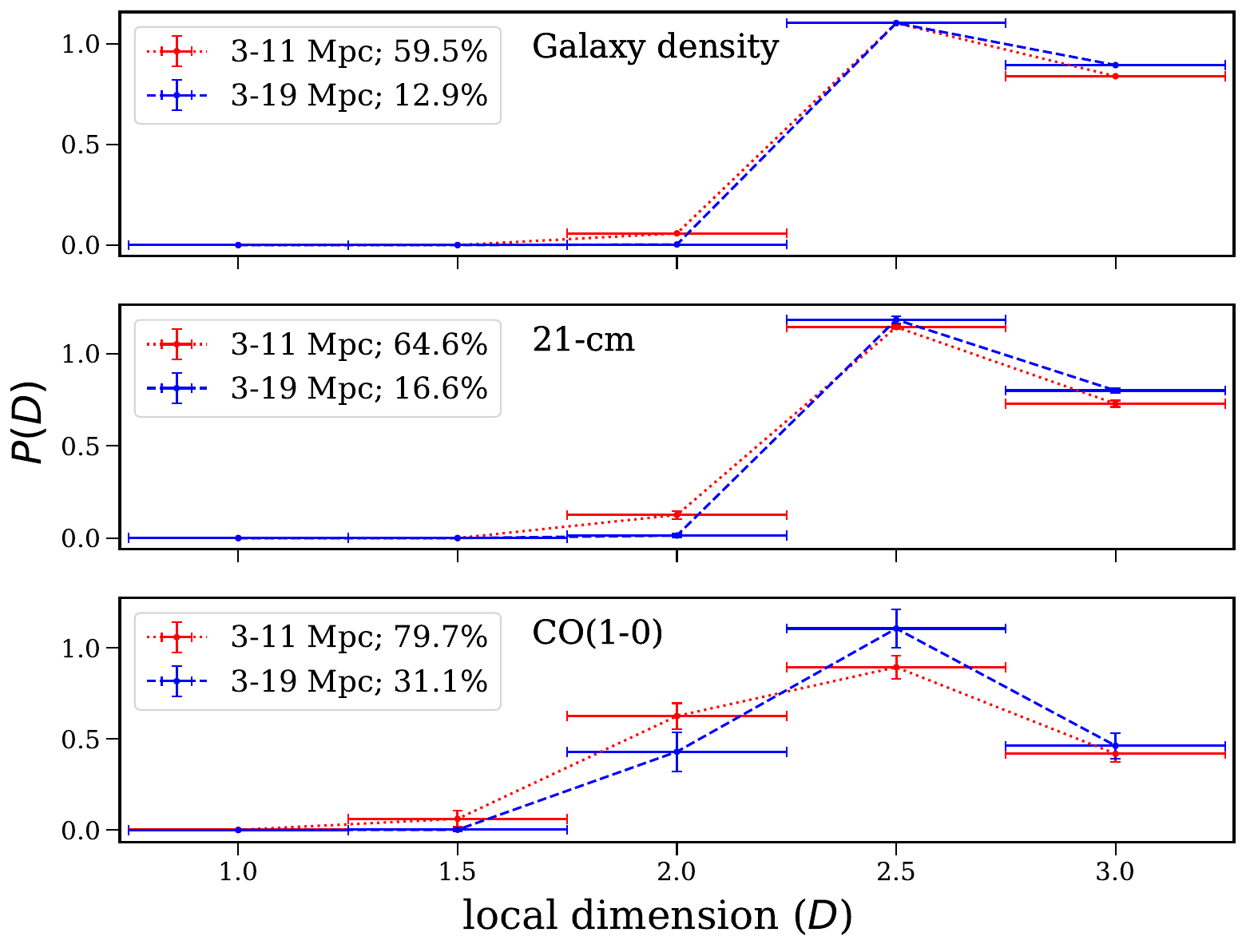}
    \caption{Same as Figure \ref{fig:locdim_signal_real}, but with galaxy positions in redshift space after applying RSD in the Kaiser regime.}
    \label{fig:locdim_signal_RSD}
\end{figure}

In redshift space, the distribution shifts towards lower $D$ in each map as shown in Figure \ref{fig:locdim_signal_RSD}. The galaxy density and $21$-cm maps retain a distribution of local dimensions similar to each other; however, the peak shifts to I2. As in real space, the CO($1-0$) map shows a morphology shifted towards lower $D$-values across length scales. As the Kaiser effect leads to increased clustering along the line of sight, structures become squashed and appear more filamentary perpendicular to the line of sight. This leads to the shift in the local dimensions towards lower (more filamentary) values. Henceforth, all maps used for morphological analysis will be in redshift space unless specified otherwise. The suppression of CO($1-0$) emission contribution from the C3 class and the increase in the relative contribution from I1 and C2 might result from a lower abundance of star-forming galaxies in void-like environments than inside the web.

The distribution of local dimensions is expected to depend on the spatial resolution of the maps. Filamentary features on small scales, which can be resolved at a high resolution, get smoothed out on going to a coarser resolution. This is shown in Figure \ref{fig:resolution_comparison} for CO($1-0$) maps at four different spatial resolutions: $\delta s \simeq 0.25,\, 0.5,\, 0.75, {\rm and \,} 1$ Mpc. With increasing spatial resolution, the distribution across length scales shifts towards lower $D$-values.

\subsubsection{Thermal noise-contaminated maps}
So far, we have seen that the local dimension distributions show different signatures for the $21$-cm and CO($1-0$) maps. We now study the detectability of these features in SKA-Mid observations in the presence of thermal noise. We find that at least $5000$ hours per pointing of SKA1-Mid observation are required to achieve an appreciable SNR for the $21$-cm map at the required resolution. On the other hand, the CO($1-0$) map attains signal dominance after $100$ hours of SKA2-Mid observation per pointing.

\begin{figure}[htbp]
    \centering
    \includegraphics[width=0.7\linewidth]{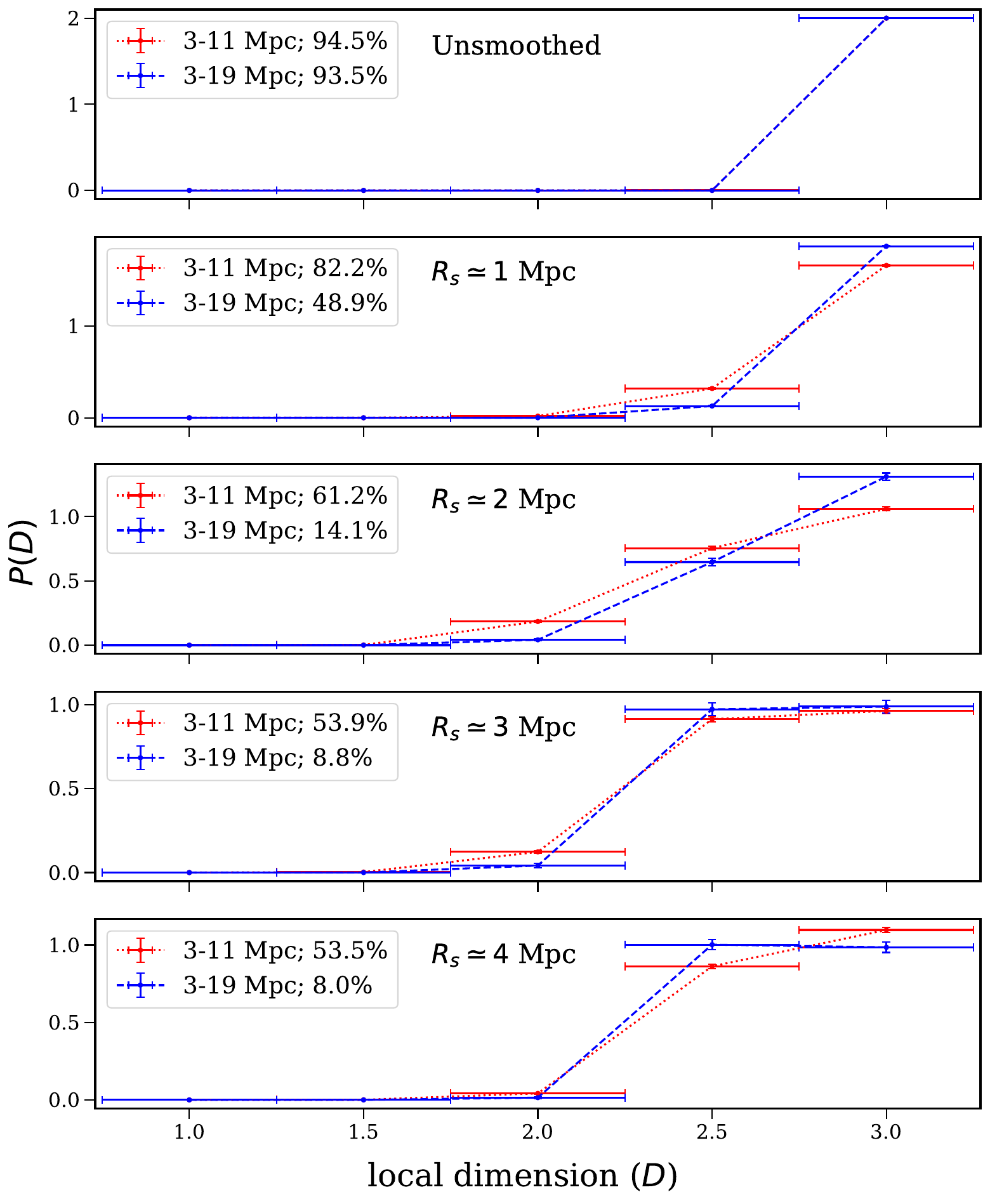}
    \caption{Distribution of local dimensions recovered from the $21$-cm map contaminated with Gaussian random thermal noise corresponding to $5000$ hours per pointing of SKA1-Mid observations. The threshold is set at $1\sigma$ (noise standard deviation). From top to bottom, we show the distribution for the unsmoothed map and after de-noising the map with Gaussian smoothing kernels of sizes $R_{\rm s} \simeq 1,\, 2,\, 3,\, {\rm and}\, 4$ Mpc. The error bars on $P(D)$ are $3\sigma$ uncertainties over $100$ noise realisations.}
    \label{fig:HI_noisy_1sigma}
\end{figure}

We show the results upon adding Gaussian random thermal noise for $5000$ hours per pointing of SKA1-Mid observations in the AA4 configuration for the $21$-cm map in Figure \ref{fig:HI_noisy_1sigma}. The normalised fractions $P(D)$ are estimated by averaging over $100$ noise realisations. On applying a $1\sigma$ threshold, the local dimension distribution gets washed out, and all sampled cells occupy C3. This is due to a uniform distribution of bright cells resulting from noise dominance even above $1\sigma$. A similar trend is seen with a $2\sigma$ threshold as well. We apply smoothing filters of sizes $R_ {\rm s} = m \, \delta s$ where $m=1,\,2,\,3,\,4$ and $\delta s = 0.87$ Mpc. On increasing the smoothing scale, the distribution starts to appear different from the unsmoothed case. The contribution from I2 and C2 classes increases, and at $R_ {\rm s} \simeq 4$ Mpc, the distribution starts to qualitatively resemble that of the uncontaminated signal in the middle panel of Figure \ref{fig:locdim_signal_RSD}. However, the exact nature of the distribution is markedly different. Additionally, the percentage of classifiable cells decreases with increasing smoothing scale.

\begin{figure}[htbp]
    \centering
    \includegraphics[width=0.65\linewidth]{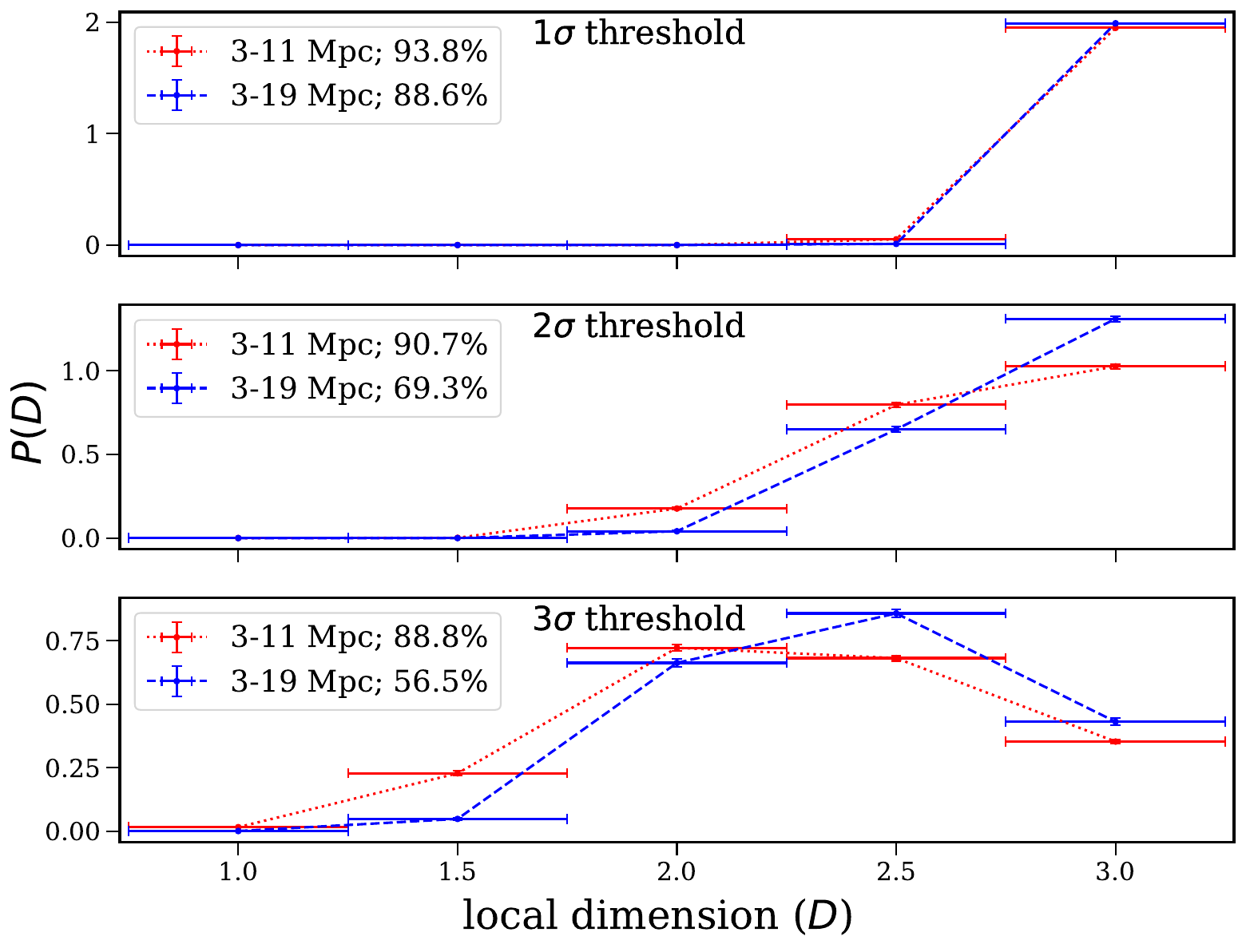}
    \caption{Distribution of local dimensions recovered from the CO($1-0$) map contaminated with Gaussian random thermal noise corresponding to $100$ hours per pointing of SKA2-Mid observations. From top to bottom, we show the distribution for $1\sigma$, $2\sigma$, and $3\sigma$ thresholds, where $\sigma$ is the standard deviation of noise. The error bars on $P(D)$ are $3\sigma$ uncertainties over $100$ noise realisations.}
    \label{fig:CO_noisy}
\end{figure}

Figure \ref{fig:CO_noisy} shows the distribution of local dimensions of the CO($1-0$) map after adding thermal noise corresponding to $100$ hours of SKA2-Mid observations per pointing, assuming an AA4 configuration of dishes. As in the case of the $21$-cm map, the distribution is confined to C3 when a $1\sigma$ threshold is applied. However, it changes significantly when a $2\sigma$ threshold is applied, with increased contributions from C2 and I2. The percentage of classifiable cells is high, with $\sim 90\%$ and $\sim 70\%$ of cells classifiable in the $3-11$ and $3-19$ Mpc length scales, respectively. Most importantly, on applying a $3\sigma$ threshold, the distribution better resembles that of the noiseless CO($1-0$) map (bottom panel of Figure \ref{fig:locdim_signal_RSD}), and the percentage of classifiable cells also remains high. The recovered distribution is shifted towards lower $D$-values, but this behaviour is consistent in both the length scales.
This indicates that in the presence of thermal noise only, the morphology of CO($1-0$) intensity maps on sub-Mpc length scales at $z \simeq 1.4$ can be recovered with a minimal bias by SKA2-Mid by means of the local dimension.

\subsubsection{Interloper-contaminated maps}

\begin{figure}[htbp]
    \centering
    \includegraphics[width=\textwidth]{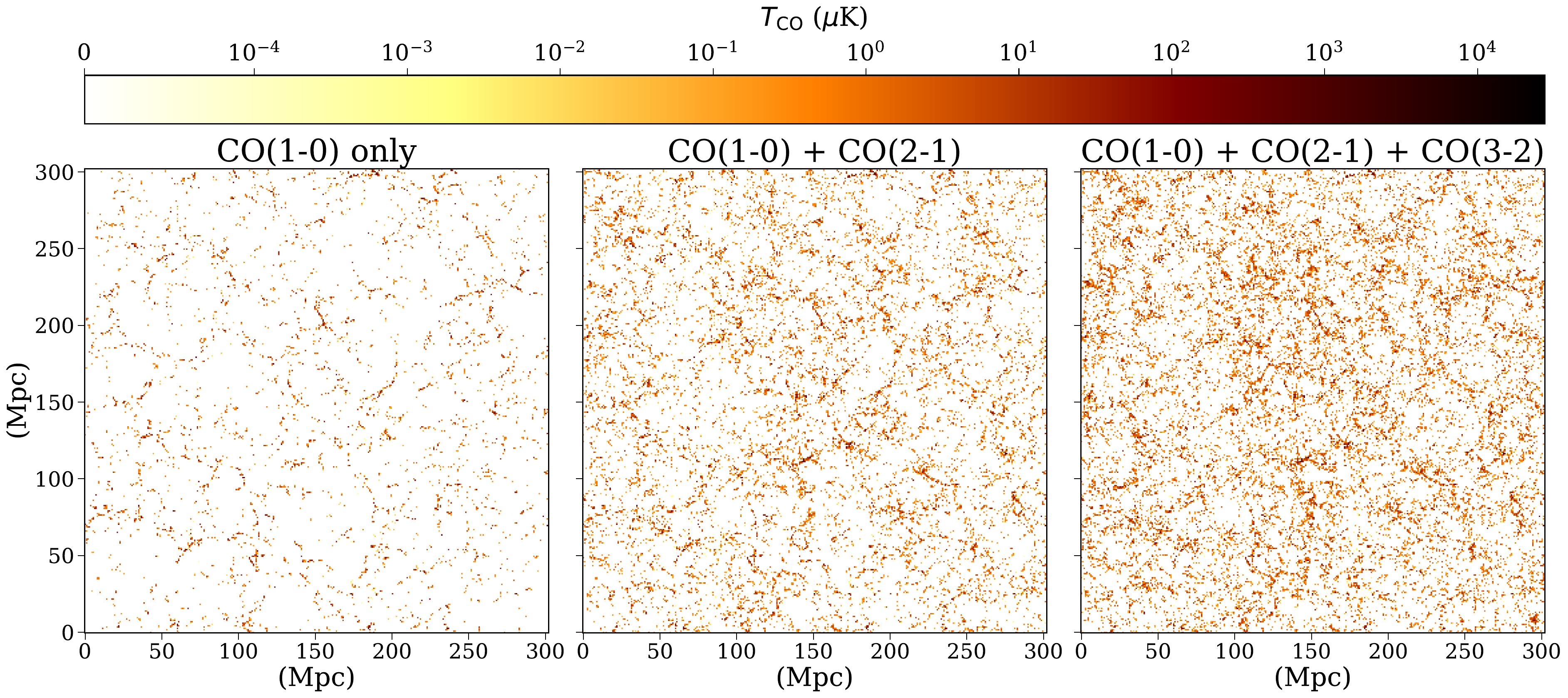}
    \caption{Left: Slice of the CO($1-0$) brightness temperature lightcone centred at $z=1.41$ with spatial resolution $\delta x \simeq 0.87$ Mpc; the same slice of the lightcone contaminated by CO($2-1$) from $z=3.82$ only (middle) and the sum of CO($2-1$) from $z=3.82$ and CO($3-2$) from $z=6.23$ (right). The horizontal axis is the line of sight with redshift increasing from left to right.}
    \label{fig:contam_maps}
\end{figure}

Figure \ref{fig:contam_maps} shows slices parallel to the line of sight of the CO($1-0$) lightcone centred at $z=1.41$ (left), the same lightcone contaminated by the two strongest interlopers: The CO($2-1$) emission from $z=3.82$ only (middle), and the resultant of CO($2-1$) at $z=3.82$ and CO($3-2$) from $z=6.23$ (right). The maps clearly exhibit washing out of the target structures by the interlopers.

\begin{figure}[htbp]
    \centering
    \includegraphics[width=0.7\linewidth]{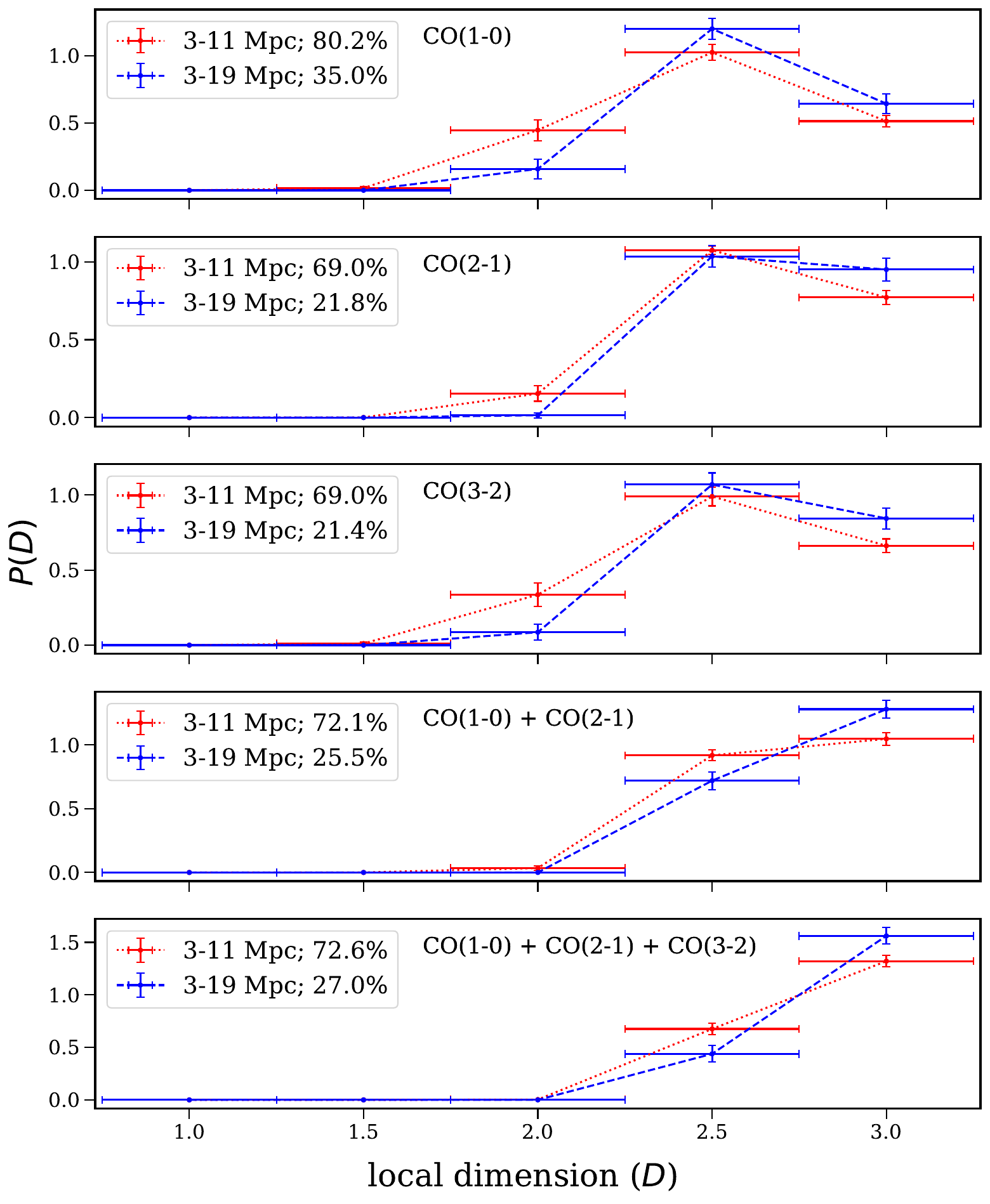}
    \caption{Distribution of local dimensions of $10^5$ randomly sampled bright cells in the (from top to bottom) (i) target CO($1-0$) lightcone centred at $z=1.41$, (ii) interloper CO($2-1$) lightcone centred at $z=3.82$, (iii) interloper CO($3-2$) lightcone centred at $z=6.23$, (iv) CO($1-0$) contaminated by CO($2-1$) only, and (v) CO($1-0$) contaminated by CO($2-1$) and CO($3-2$). The $1\sigma$ error bars denote standard deviation in $P(D)$ across $27 \, (100 \, {\rm Mpc})^3$ subcubes, corresponding to a realistic SKA-Mid sub-survey for $10^4$ randomly sampled bright cells.}
    \label{fig:interloper}
\end{figure}

The distributions of local dimensions of the target CO($1-0$) lightcone and the lightcones of the individual interloper lines centred at the redshifts from which they contribute are shown in the first three panels of Figure \ref{fig:interloper}. The distribution for the CO($1-0$) lightcone looks almost identical to that of the coeval CO($1-0$) intensity map as shown in the bottom panel of Figure \ref{fig:locdim_signal_real}. The distributions for both the interloper-only lightcones are almost identical, with the percentages of classifiable centres for the different length scales also agreeing very closely. The distribution differs from that of the target lightcone noticeably, as the contribution from the C3 class increases across length scales. This might be a result of a smoother and less clustered LSS at higher redshifts.

The bottom two panels of Figure \ref{fig:interloper} show the local dimension distribution for the resultant maps after accounting for the interloper contribution from CO($2-1$) only and the aggregate of CO($2-1$) and CO($3-2$). The original distribution for the CO($1-0$) lightcone is distorted even after accounting for the contamination by CO($2-1$) only, as the contribution from the C3 class increases sharply and the peak shifts to C3 for each length scale probed. The contribution from the C2 class also almost vanishes at each length scale, and any sheet-like behaviour becomes undetectable. Adding the contribution from CO($3-2$) further skews the distribution towards C3, with the relative contribution from I2 dropping and that from C2 vanishing completely. The percentages of classifiable cells increase marginally at all scales compared to the panel above, but are lower than those in the target signal. This indicates that the structure appears increasingly homogeneous as the interloper contributions are accounted for, and that interloper signals must be modelled separately to isolate the morphology of the intensity maps of the target signal.

\section{Summary and conclusion}   
\label{sec:conclusions}

Line intensity mapping is a useful complementary probe to galaxy surveys to characterise the large-scale structure of the universe, which forms the cosmic web consisting of voids, sheets, filaments, and nodes. SKA1-Mid will produce tomographic intensity maps of the $21$-cm line up to $z \simeq 3$, and in phase 2 (SKA2-Mid), both the $21$-cm and CO($1-0$) lines may be observed in an overlapping redshift range of $1.4 \lesssim z \lesssim 3$. To trace the cosmic web, we targeted the atomic (HI) and molecular (H$_2$) phases of hydrogen --- the most abundant element in the universe. The $21$-cm line traces HI gas, whereas molecular gas is traced by the CO($1-0$) rotational line emission. We made one of the first attempts to characterise the LSS morphology directly with LIM images in the post-reionisation universe, where HI gas is confined to galaxies. This is an alternative strategy to Fourier domain summary statistics, such as the power spectrum, and preserves the non-Gaussian and morphological information that is lost when the information of a field is compressed into such statistics.

We post-processed the galaxy catalogues from the TNG300-1 box of the gravo - magnetohydrodynamical simulation IllustrisTNG using the Line Intensity Mapping Tool (\texttt{LIMiT}). We produced galaxy mass density, and $21$-cm and CO($1-0$) line intensity maps by extracting the total galaxy masses, gas masses, hydrogen fractions, and SFRs of galaxies. We also implemented the effect of RSD using the peculiar velocities of galaxies along the line of sight.

Simulations of mock observations with thermal noise for the $21$-cm and CO($1-0$) maps followed the specifications of the AA4 configuration for SKA1-Mid and SKA2-Mid (spectral bandwidth upgraded), respectively. We simulated Gaussian random noise for the above telescopes and computed the SNRs at different redshifts for $5000$ hours of observation per pointing for the $21$-cm signal and $100$ hours of observation per pointing for the CO($1-0$) signal. We found that an optimal SNR, where the morphologies of both line emissions can be analysed, is attained at $z = 1.41$. We also found that a balance between a coarse enough resolution, where the noise is minimised, and a fine enough resolution, where sub-Mpc scale structure is resolved, is achieved by averaging over eight frequency channels of SKA-Mid. This yields an effective spectral resolution of $107.52$ kHz, corresponding to a comoving spatial resolution of $0.87$ Mpc at $z = 1.41$.

A single pointing of SKA-Mid can survey $5.56 \, {\rm deg}^2$ at a central frequency of $589$ MHz (redshifted frequency of the 21-cm line from $z=1.41$). This corresponds to a comoving volume of $(177 \, {\rm Mpc})^3$ for a spectral bandwidth of $\simeq 22$ MHz. Therefore, a cosmological volume can be surveyed using the 21-cm line by observing at a single pointing for $5000$ hours. However, due to the frequency dependence of the primary beam width, the field of view of the same array for the CO($1-0$) line is limited to $\simeq 92 \, {\rm arcmin}^2$. This is considering a phased array system of feeds offset from each other in such a way that it increases the field of view of the array by $\sim 30$ times. Such a strategy has been employed by the Australian SKA Pathfinder (ASKAP) \citep{Hotan_ASKAP_2021}, and a similar approach has been used by the COMAP Collaboration to observe the CO LIM signal in the post-reionisation universe \citep{lunde_comap_2024}. CO($1-0$) intensity maps formed using approximately 70 non-overlapping pointings of the array at central frequency $48$ GHz with spectral bandwidth $\simeq 1$ GHz must be mosaicked to survey a comoving volume of $(100 \, {\rm Mpc})^3$, which is the minimum cosmological volume that can be assumed to show statistical homogeneity and isotropy. The resulting 21-cm and CO($1-0$) sub-surveys will effectively require $5000$ and $7000$ total observing hours, respectively.

We studied the percolation transition of the galaxy mass density map, as well as the $21$-cm and CO($1-0$) line intensity maps, using the largest cluster statistic (LCS) and filling factor (FF) by iteratively coarse-graining the maps. The galaxy density map showed the slowest percolation transition, followed by the $21$-cm and CO($1-0$) maps. As the FF could not be independently varied, the LCS values could not be compared at the same FFs. Nevertheless, the CO($1-0$) map appeared to percolate faster than the $21$-cm map. It was demonstrated in \citep{bharadwaj_evidence_2000} that a faster percolation transition indicates a higher degree of overall filamentarity of a field. Since a significant difference in the FFs at which the percolation transition occurs for the different maps could not be established, a higher degree of filamentarity of the CO($1-0$) map is not conclusive. However, the three maps tracing different percolation curves, despite being biased tracers of the same matter distribution, indicate that they have differing morphologies.

To confirm whether CO($1-0$) emission in the intensity map originates from biased locations in the cosmic web, we computed local dimensions of cells lying in overdensities in the galaxy distribution and hot spots in the $21$-cm and CO($1-0$) line intensity maps. We then determined what fractions of these cells belong to the different structural elements in the cosmic web. Out of the grid cells in the maps that were classifiable using the local dimension, the largest fraction belonged to void-like environments, followed by regions intermediate between sheets and voids. This can be expected as voids account for $\sim95\%$ of the volume of the universe. The $21$-cm map showed a similar behaviour, along with an increase in the fraction of classifiable cells. This indicates that the $21$-cm line emission traces the galaxy distribution almost exactly, with slight differences arising due to gas-depleted galaxies. However, the classifiable bright cells in the CO($1-0$) map have the largest contribution from environments intermediate between sheets and voids, and the relative contribution from void-like regions is suppressed.

Even the absolute fraction of cells in sheet-like environments, which increases sharply in the CO($1-0$) map compared to the $21$-cm map, comes at the expense of voids, as the fraction in the intermediate environment also increases. Since the CO($1-0$) luminosities of galaxies are entirely SFR-dependent in our model, this indicates quenching of star formation in such environments. This disagrees with the findings of \citep{kraljic_galaxy_2018, pandey_exploring_2025}, where voids have high fractions of blue and star-forming galaxies.

The local dimension distribution shifts towards lower $D$-values for maps in redshift space due to increased clustering of structure along the line of sight. This makes structures appear more filamentary than they actually are. The negligible contribution of filaments in each case is a combination of two factors: Firstly, our resolution element is large, and hence filamentary features might go unresolved. Secondly, we computed the normalised fraction of bright cells that lie within each structural element, which is related to the volume it occupies. As filaments occupy a lesser volume than sheets and voids, the fractions are extremely low.

We found that the local dimension distribution depends on the spatial resolution of the maps, and filamentary features start to get resolved with finer resolutions. We find that the cosmic variance introduced in the distribution of local dimensions due to sub-surveys of the larger $(300 \, {\rm Mpc})^3$ volume is small compared to the actual distributions obtained from the larger volume.

We studied the prospects of recovering the distribution of local dimensions from $21$-cm maps contaminated by thermal noise corresponding to $5000$ hours of SKA1-Mid observation. On applying a $1\sigma$ noise threshold to the noise-contaminated map, the ground truth morphology was found to be completely destroyed and unrecoverable. We smoothed the map with Gaussian kernels of varying sizes to de-noise the map and found that at kernel sizes of $\sim 3-4$ Mpc, the nature of the distribution starts to get recovered. However, the exact features of the distribution differ from those of the uncontaminated map. We conclude that the morphology of $21$-cm intensity maps at sub-Mpc length scales at $z \simeq 1.4$ cannot be recovered accurately even with $5000$ hours of SKA1-Mid observation.

The recovery prospects were found to be better for CO($1-0$) intensity maps contaminated by thermal noise only with SKA2-Mid. On adding thermal noise corresponding to $100$ hours per pointing (7000 total observing hours) of SKA2-Mid observation and applying a $1\sigma$ noise threshold, the morphology could not be recovered as in the case of the $21$-cm map. However, applying a $3\sigma$ noise threshold yielded the ground truth morphology of the CO($1-0$) map more closely, without the need for de-noising.

The most important limiting factor in the detectability of CO($1-0$) intensity maps was found to be the contamination from interlopers. Even when the contribution from the closest interloper line CO($2-1$) was accounted for, the inferred morphology was very different from the true morphology of the CO($1-0$) intensity map. The morphology resembled a homogeneous distribution even more when the contamination from CO($3-2$) was added. This implies that, given an observed CO intensity map with the foregrounds and other systematic effects (out of the scope of this study) modelled perfectly, modelling the tomographic intensity maps of the interloper lines CO($2-1$) and CO($3-2$) is necessary to recover the target CO($1-0$) intensity map and its morphology.

The intensity map simulations in this work assumed a one-to-one relation between the HI masses of galaxies and their $21$-cm brightness temperatures, and the SFRs and their CO($1-0$) luminosities, without accounting for the line luminosity scatter in these relations. Furthermore, a constant neutral fraction of hydrogen in galaxies was assumed. Another simplifying assumption was made about the comoving grid resolutions being the same along the line of sight and on the sky plane, which is not generally true for observed intensity maps. The
contamination by residual foregrounds, ionospheric distortions, atmospheric absorption, etc. were disregarded, which are expected to have a significant impact on signal detectability. We plan to make the simulations more realistic by implementing the above effects in future work.


In this work, we have explored the local dimension as an image-based morphological measure to describe line intensity maps. It can be employed as a higher-order summary statistic for the estimation of astrophysical and cosmological parameters in a Bayesian inference framework, as it is sensitive to morphological features. However, despite being a simple exponent to compute, the local dimension is highly resolution-dependent and relies on the convergence of a power law fit to be defined. The eigenvalues of the tidal tensor \citep{hahn_properties_2007, foreroromero_dynamical_2009, aycoberry_theoretical_2023} can be a useful alternative method to characterise the contribution of cosmic web environments to intensity maps. This has the potential to be a more robust morphological measure, overcoming the above-mentioned drawbacks of the local dimension.


\acknowledgments
We would like to thank the anonymous reviewer for their comments and suggestions, which have strengthened the analysis presented in this work. MMD, SM, and AD would like to thank the Science and Engineering Research Board (SERB) and the Department of Science and Technology (DST), Government of India, for financial support through Core Research Grant No. CRG/2021/004025 titled “Observing the Cosmic Dawn in Multicolour using Next Generation Telescopes”. MMD and SM also thank Somnath Bharadwaj for helpful suggestions on percolation and local dimension analyses. MMD also thanks Leon Noble for useful discussions. CSM would like to acknowledge financial support from the Council of Scientific and Industrial Research (CSIR) via a CSIR-SRF Fellowship (Grant No. 09/1022(0080)/2019-EMR-I) and from the ARCO Prize Fellowship. SKP acknowledges financial support from the Department of Science and Technology, Government of India, through the INSPIRE Fellowship [IF200312]. SD would like to acknowledge the Cambridge Trust and Isaac Newton Studentship for funding his PhD. SB acknowledges the funding provided by the Alexander von Humboldt Foundation and the Deutsche Forschungsgemeinschaft (DFG, German Research Foundation) under Germany's Excellence Strategy -- EXC-2094 -- 390783311. SB thanks Varun Sahni, Prakash Sarkar, and Santanu Das for their contributions to the development of \texttt{SURFGEN2}.

\bibliographystyle{JHEP}
\bibliography{bibliography}

\end{document}